# Scale-bridging dislocation plasticity in MgO at room temperature

Jiawen Zhang[1], Zhangtao Li[2], Yuwei Zhang[3], Hendrik Holz[3], James P. Best[3], Oliver Preuß[4], Zhenyong Chen[5], Yinan Cui[2*], Xufei Fang[6*], Wenjun Lu[1*]

[1]Department of Mechanical and Energy Engineering, Southern University of Science and Technology, Shenzhen 518055, China

[2]Department of Engineering Mechanics, Tsinghua University, Beijing 100084, China

[3]Max Planck Institute for Sustainable Materials, Düsseldorf 40237, Germany

[4]Department of Materials and Earth Sciences, Technical University of Darmstadt, Darmstadt 64287, Germany

[5]Bay Area Center for Electron Microscopy, Songshan Lake Materials Laboratory, Dongguan 523808, China

[6]Institute for Applied Materials, Karlsruhe Institute of Technology, Karlsruhe 76131, Germany

*Corresponding authors: cyn@mail.tsinghua.edu.cn; xufei.fang@kit.edu; luwj@sustech.edu.cn

## Abstract

Dislocations in ceramics have recently gained renewed research interest, in contrast to the traditional belief that ceramics are inherently brittle. Understanding dislocation mechanics in representative oxides is beneficial for effective dislocation engineering. Here, we use MgO single crystals with mechanically seeded dislocation densities from ~$10^{12}$ to ~$10^{15}$ $m^{-2}$ to investigate the mechanical behavior such as yield and fracture. Micro-pillar compression tests reveal a dislocation density dependent yield strength, mediated by the varying dominating dislocation mechanisms from nucleation to multiplication/motion. In situ TEM compression measurements highlight the dislocation-seeded samples can achieve a much-improved compressive plastic strain beyond ~70%, with a high yield strength of ~2.35 GPa (diameter of ~400 nm), indicating size effect. Complementary bulk compression tests, along with digital image correlation (DIC), demonstrate a consistent dislocation-mediated deformation and a notable size effect, with bulk samples exhibiting much reduced yield strength (~120 MPa) compared to the nano-/micro-pillars. Using three-dimensional Discrete Dislocation Dynamics (3D-DDD) simulation, we further qualitatively analyze the collective dislocation activities (slip events) and work hardening during compression. This study provides new insights into dislocation-mediated plasticity in MgO, across different length scales, by systematically tuning dislocation density.

## 1. Introduction

The inherent brittleness of ceramic materials at room temperature has long restricted their applications as structural materials, despite their exceptional hardness and thermal stability. Unlike metals, where plastic deformation is primarily governed by dislocations, ceramics are characterized by strong ionic and covalent bonds that hinder dislocation nucleation, mobility, and multiplication (De Leon et al., 2015). This results in poor plasticity and low fracture toughness in most ceramics, with fracture often occurring before dislocations are activated (Jia et al., 2020; Li et al., 2019). Research on amorphous ceramics and ceramic composites has shown improved mechanical properties, including enhanced fracture toughness, strength, and ductility. For instance, diamond-like carbon demonstrates self-toughening with dimension down to sub-micron scales (Shin and Jang, 2020). Nanostructure amorphous $Al_2O_3$-$ZrO_2$ ceramic exhibits a medium-temperature (600 °C) plasticity of ~15% (Wang et al., 2024). Besides, Fe–SiOC amorphous ceramic composites display ~70% compressive plasticity with 34 at.% Fe deposited (Ming et al., 2020). Micro-$B_4C$/nano-MgO + CNTs-reinforced aluminum alloy composite displayed enhanced combination of strength and ductility by stacking faults and dislocation enrichment (Saba et al., 2025). Amorphous $B_4C$ reinforced copper composites exhibit improved ductility and strength(Ye et al., 2025). Stress-induced martensite phase transition from tetragonal $ZrO_2$ to monoclinic $ZrO_2$ resulted in the toughening $ZrO_2$/MgO coating on Mg alloy (Qian et al., 2023). Nevertheless, both amorphization and material compositing are restricted to only a few ceramic systems and are not suitable for most ceramics. By contrast, recent studies have revealed that pre-engineered or mechanically seeded dislocations can significantly alter the mechanical behavior of ceramics intrinsically in an effective way (Fang et al., 2024; Shen et al., 2024; Zhang et al., 2025). By engineering dislocations within ceramic materials, they can exhibit plasticity similar to that seen in metals, opening new avenues for improving the performance of ceramic materials. In particular, the ability to control dislocation densities has emerged as a promising strategy for enhancing the mechanical properties of ceramics (Fang, 2023).

Recently, the approach of engineering pre-existing dislocations on the room-temperature plasticity in ceramics has been proposed and studied. The central idea is to circumvent dislocation nucleation while promoting dislocation multiplication and motion to achieve larger degree of plastic deformation. In metals, plastic deformation is governed by the dislocation nucleation and motion (Cappola et al., 2024;

Kishida et al., 2023), while the same is true for ceramics. For instance, Shen et al. reported that high-temperature pre-loading induced dislocations and stacking faults lead to ~10% and ~7% plastic strain for micro-pillar compression in $TiO_2$ and $Al_2O_3$, respectively (Shen et al., 2024). While high-temperature deformation has been widely employed to facilitate dislocation motion in ceramics (Heuer, 2005; Nishigaki et al., 1991; Shen et al., 2024), techniques such as room-temperature nanoindentation and cyclic indentation or scratching have proven efficient in generating dislocations by mechanical deformation in various oxide ceramics, including MgO (Preuß et al., 2024; Tromas et al., 2006), $SrTiO_3$ (Fang et al., 2021; Zhang et al., 2025), and $KNbO_3$ (Preuß et al., 2023). For instance, dislocation enrichment by surface grinding in single-crystal $SrTiO_3$ leads to the increase of compression strain over ~30% (Fang et al., 2024). Furthermore, mechanical imprinting of various dislocation densities by cyclic scratching have been recently demonstrated to effectively impact the yield strength and plastic strain in $SrTiO_3$ (Zhang et al., 2025). Although these mechanically seeded dislocations have been shown to enhance the compression plasticity and fracture toughness of ceramics (Fang et al., 2024; Moon and Saka, 2000; Preuß et al., 2023), their exact contribution to the deformation behavior remains an open question except for $SrTiO_3$ (Fang et al., 2024; Zhang et al., 2025), which also raises the question for the general applicability of the approach using mechanically seeded dislocations. As most recently reviewed by Frisch et al. (Frisch et al., 2025) there have been 44 ceramic compounds reported so far exhibiting room-temperature dislocation plasticity at meso/macro-scale, beyond the nano/micro-scale plastic deformation with confined volume. This naturally leads to the following questions to be addressed in this study: Can these findings in $SrTiO_3$ be generalized to other oxide ceramics? How do these mechanically seeded dislocations impact the length-scale dependent mechanical properties and dislocation behavior?

This study aims to address the impact of variable dislocation densities on the mechanical properties of magnesium oxide (MgO) across nano/micro- to macro-scale. MgO, with its simple rock-salt structure, has been a model material for investigating the dislocation-mediated plasticity in oxide ceramics. Early studies by Gorum et al. demonstrated that MgO could bend without cracking and displayed plastic compression strain at room temperature at bulk scale (Gorum et al., 1958). Other bulk compression tests investigated by Argon et al., Orowan et al., and Stoke et al. found that MgO single crystal could deform plastically by room temperature bulk compression, accompanied by the generation of slip traces (Argon and Orowan, 1964; Stokes et al., 1961, 1959). Moreover, MgO could even plastically deform at room temperature tensile test (~5% total strain) with pre-existing dislocations induced by surface polishing (Stokes, 1963). Later studies, particularly involving nanoindentation and *in situ* transmission electron microscopy (TEM), revealed that dislocations could be nucleated and activated at room temperature in MgO (Issa et al., 2021, 2015; Tromas et al., 2006). Simulation work by Cordier et al. demonstrated that MgO could deform in athermal regime, resulting in a very weak phase (Cordier et al., 2012). Although dislocation-driven plasticity in MgO at room temperature is understood to be dominated by dislocation glide and nucleation (Amodeo et al., 2018), the research gap remains that neither tunable dislocation densities nor a systematic scale-bridging study on dislocation plasticity in MgO have been achieved.

Here, we employ cyclic scratching to tune dislocation densities in (001) MgO single crystals, followed by nano/micro-pillar compression tests to evaluate their mechanical properties. At the macro-scale, bulk compression tests, combined with digital image correlation (DIC), are used to examine the strain field and fracture behavior. These experimental techniques are complemented by three-dimensional Discrete Dislocation Dynamics (3D-DDD) simulations, which allow us to examine the collective dislocation behavior and its impact on ceramic plasticity at the nano/micro-scales. Unlike previous work on $SrTiO_3$ that lacks simulation insights (Zhang et al., 2025), this study combines multi-scale experiments and simulations for more complete investigation of dislocation-mediated plasticity in MgO with different pre-engineered dislocation densities. Focus on key mechanisms such as dislocation multiplication, dislocation cell formation, and the role of size effects in deformation, and reveal the different mechanical responses in rock-salt MgO.

## 2. Methods

### 2.1. Material selection

MgO single crystals with dimensions of 5 × 5 × 1 $mm^3$ (Alineason Materials Technology GmbH, Frankfurt am Main, Germany) are used for scanning probe microscope (SPM) examination. The 5 × 5 $mm^2$ surface with an orientation of (001) was selected for all experimental tests. The sample surfaces were initially polished using SiC abrasive paper down to a grit size of 6 μm, followed by diamond polishing with suspensions of 3 μm, 1 μm, and 0.25 μm. To achieve a final surface roughness of less than 2 nm, vibration polishing with OPS suspension was conducted for about 16 h, as confirmed by the scanning probe microscopy (SPM) image in Figure S1. This meticulous polishing procedure also minimized the presence of mechanically induced dislocations by grinding and polishing in the near-surface region. For room-temperature pillar compression and bulk compression measurements, (001) MgO single crystal samples with dimensions of 6 × 3 × 3 $mm^3$ (6 mm in height) were obtained from Hefei Ruijing Optoelectronics Technology Co., Ltd. (Anhui, China). All surfaces of these samples were pre-polished by the manufacturer to achieve a surface roughness of less than 2 nm, as also shown in Figure S1. This ensured the samples are suitable for high-precision mechanical testing.

**2.2. Dislocation imprinting via cyclic scratching**

To achieve tunable dislocation densities, cyclic scratching was conducted at room temperature (25 °C) using a 2.5 mm diameter hardened steel Brinell indenter. The indenter was lubricated with methyl silicone oil to minimize wear and prevent cracking. Scratching parameters were optimized based on prior research (Preuß et al., 2024) with the following settings: a scratching speed of 0.5 mm/s, a normal load of 1.5 kg, and a scratching distance of about 5 mm along the [001] direction. These optimizations were performed using a universal hardness tester (Finotest, Karl Frank GmbH, Weinheim, Germany). To achieve varying dislocation densities, scratching passes were set to 1× and 10× (× represents the number of passes), with the latter involving bi-directional scratching. This approach allowed for precise control over dislocation density, enabling subsequent analysis of its effects on the mechanical properties of the samples.

**2.3. Mechanical testing from nano-, micro-, to macro-scale**

The (001) MgO micro-pillars, with varying dislocation densities (0×, 1×, and 10× scratching), were prepared using a dual-beam focus ion beam (FIB) microscope (Helios Nanolab 600i, FEI, Hillsboro, USA) equipped with a scanning electron microscopy (SEM). Each micro-pillar was fabricated from the middle of the wear tracks, with diameters of 400 nm, 500 nm, 1 μm, 3 μm, and 5 μm. Room-temperature compression tests were conducted using a nanoindentation instrument (Hysitron TI950, Bruker, USA) with a 5 μm flat punch diamond indenter at a constant strain rate of $1 \times 10^{-3}$ $s^{-1}$. To assess the electron beam effects, pillars with 1 μm and 3 μm diameter were tested outside the SEM, while the 5 μm diameter pillars were tested both in- and outside the SEM. The preset deformation strain was 25%-40% for all samples. For *in situ* micro-pillar compression tests, larger micro-pillars with a diameter of 5 μm were milled using the Ga FIB/SEM system (Thermo Fisher Helios 5 CX, USA). The micro-pillars were tested using an *in situ* micromechanical testing system (Alemnis AG, Switzerland) inside a SEM (Zeiss Gemini 500, German) using a flat punch diamond indenter with a diameter of 10 μm. The compression strain rate for *in situ* tests was set at $2.5 \times 10^{-4}$ $s^{-1}$, and all conditions were repeated five times. *In situ* transmission electron microscopy (TEM: Thermo Fisher, Talos F200X G2, USA) compression tests were performed on a PI95 Picoindenter (Hysitron, USA) using displacement-controlled mode and a constant strain rate of $1 \times 10^{-3}$ $s^{-1}$.

Room temperature bulk compression tests on MgO samples were conducted along the <001> direction using an MTS E45 mechanical testing machine. Prior to testing, two polycrystalline $Al_2O_3$ plates (~1 mm thick) were placed on the top and the bottom surfaces of the MgO sample to ensure smooth and uniform contact. The compression strain rate was maintained at $1.5 \times 10^{-4}$ $s^{-1}$. During compression, *in situ* deformation images were captured at a rate of 500 ms per frame using VIC-GaugeB2D software (Correlated Solutions, Inc.). A DIC technique was employed to measure displacements and strain fields throughout the uniaxial compression tests. This setup utilized a high-resolution camera (3000 × 4000 pixels) synchronized with the tensile testing instrument. For DIC measurements, the strain range is 0-26.8%, the unit is engineering strain, reported in “%”, the subset size is 31 × 31 px, the step size is 7 px, the spatial resolution is 0.175 mm, and the error bar for xx is around 0.02%. To enhance imaging quality, the MgO bulk samples were coated with a layer of white paint, followed by black paint speckles dispersed on one surface to create a suitable contrast for DIC analysis.

### 2.4. Microstructure characterization

TEM was used to investigate variations in dislocation structures and densities across different scratching passes. For the as-scratched samples, TEM lamellae were extracted from the middle of the scratching tracks (1× and 10×) along the scratching direction (<001> direction) using a FIB instrument. Reference lamellae were prepared from regions distant from the scratching tracks to serve as a baseline. The initial milling and lift-out were performed at a voltage of 30 kV. The lamellae were subsequently thinned using a progressively decreasing ion beam current: 0.43 nA, 0.23 nA, 80 pA, and finally 40 pA. To minimize ion beam damage and amorphous layer formation, the final cleaning of both lamella surfaces was conducted at low energies: first at 5 kV with a 41 pA current, followed by a final polish at 2 kV with a 24 pA current. Scanning TEM (STEM) images were acquired using a Talos F200X G2 TEM operating at 200 keV. Annular dark-field STEM (ADF-STEM) images were obtained at a probe semi-convergence angle of 10.5 mrad and inner and outer semi-collection angles of 23-55 mrad. Dislocation densities of the as-scratched samples were estimated using the line intercept method based on low-magnification ADF-STEM images (Meng et al., 2021). High-resolution STEM analysis was performed on a probe aberration-corrected STEM (FEI Titan Themis) at 300 keV to visualize atomic arrangements and dislocation core structures in the as-scratched (001) MgO samples. For high-angle annular dark-field STEM (HAADF-STEM) imaging, a probe semi-convergence angle of 17 mrad was used, with inner and outer semi-collection angles ranging from 38 to 200 mrad. This comprehensive analysis enabled detailed visualization of dislocation features, providing critical insights into the effects of scratching passes on the dislocation structures in MgO.

### 2.5. Simulations setup

Three-dimensional Discrete Dislocation Dynamics (3D-DDD) is employed here to further disclose the collective dislocation behavior in MgO pillar compression, using MoDELib code (Cui et al., 2022, 2019, 2018a; Po et al., 2014; Po and Ghoniem, 2014). In this approach, dislocation lines are discretized into parameterized dislocation segments. The configurational Peach-Koehler forces are calculated according to the applied stress, the dislocation stress fields, and the image forces induced by free surface. The nodal motion equations are solved using a finite element-based methodology, as detailed in (Ghoniem et al., 2000; Ghoniem and Cui, 2020; Po et al., 2014). Additional implementation details are available in our previous work (Cui et al., 2022, 2018b).

Several researchers have studied the mobility laws of dislocations in MgO (Amodeo et al., 2018, 2011; Reali et al., 2017). The motion of screw dislocations in MgO is primarily governed by the kink-pair mechanism at room temperature as in the present case (Cordier et al., 2012), and the corresponding atomic simulation informed dislocation mobility laws have been used as follows (Kocks et al., 1975):

$$v = \begin{cases} b\dfrac{L_d}{l_c}\nu_D\dfrac{b}{l_c}\exp\left(\dfrac{-\Delta H(\tau,T)}{k_B T}\right) & \text{if } \Delta H(\tau,T) > 0 \\ \dfrac{\tau b}{B} & \text{if } \Delta H(\tau,T) \leq 0 \end{cases} \quad (1)$$

where $b$ is the magnitude of Burgers vector, $l_c$ is the critical width of kink-pairs, $L_d$ represents the length of dislocation segment, $\nu_D$ is the Debye frequency, $k_B$ is the Boltzmann constant, $T$ is the absolute temperature, and $B$ is the effective drag coefficient for dislocation glide. $\Delta H(\tau,T)$ is the activation enthalpy of kink-pair nucleation as a function of the effective resolved shear stress $\tau$ acting on the considered dislocation segment. The condition $\Delta H(\tau,T) > 0$ defines the regime of the thermal activated mechanism, where $\Delta H(\tau,T)$ is defined according to (Borde et al., 2023),

$$\Delta H(\tau,T) = \Delta H_0 \left\{ \frac{(\tau_p - \tau)}{|\tau_p - \tau|} \left[ 1 - \left( \frac{\tau}{\tau_P} \right)^p \right]^q - \frac{T}{T_a} \right\} \tag{2}$$

where $\tau_P$ is the Peierls stress, $p$ and $q$ are fitting parameters from atomic simulations. $\Delta H_0$ is enthalpy barrier for kink-pair nucleation, and $T_a$ is the athermal transition temperature. Our simulations account for edge, screw, and mixed dislocations, and automatically capture their respective roles under varying temperature and stress conditions. At low temperatures and low stresses, screw dislocations, due to their reduced mobility, tend to remain in the sample and thus play a more significant role in plastic deformation. In contrast, at elevated temperatures or high stresses—namely, when the temperature exceeds the athermal transition or the resolved shear stress surpasses the critical value such that $\Delta H(\tau,T) \leq 0$, the mobility of screw dislocations becomes comparable to that of edge dislocations, and their motion is then governed by the phonon-drag mechanism. In this regime, the slip velocity is determined by the total force acting on the dislocation segment divided by the viscous drag coefficient, corresponding to the second line of Equation (1). The parameters governing dislocation mobility, along with the material properties, are listed in Table 1.

The plastic strain rate $\dot{\boldsymbol{\varepsilon}}^p$ is calculated on the basis of the slipped area of moving dislocations (Cui et al., 2020), i.e.

$$\dot{\boldsymbol{\varepsilon}}^p = \frac{1}{\Omega\, dt} \sum_{j=1}^{N_{seg}} \left( \boldsymbol{b}^j \otimes d\boldsymbol{A}^j + d\boldsymbol{A}^j \otimes \boldsymbol{b}^j \right) \tag{3}$$

where $N_{seg}$ is the total number of the dislocation segments. $d\boldsymbol{A}$ represents the area increment during time increment $dt$ and $\boldsymbol{b}$ is the Burgers vector. The subscript $j$ indicates the $j$-th dislocation segment. $\Omega$ is the characteristic volume.

According to experimental conditions, the simulations considered a single-crystal MgO micro-pillar, compressed along the [001] orientation at room temperature, with a diameter of 0.5 μm and a height of 1 μm. Low and high initial dislocation densities $\rho_0$ were considered to study their effect on plasticity. According to the experimental results, for the low-density case, $\rho_0$ is set to $5 \times 10^{12}$ m$^{-2}$, while $\rho_0$ ~$5 \times 10^{14}$ m$^{-2}$ for the high-density case. The initial dislocation structure was generated by relaxing randomly distributed dislocations, and six <110>{110} slip systems are considered here, according to the experimental observation. The applied strain rate in all simulations is $10^3$ $s^{-1}$.

Achieving strain rates comparable to static experiments remains a considerable challenge in 3D DDD simulations due to computational cost, which typically limits DDD simulations to stain rates of $10^2$/s to $10^4$/s (Agnihotri and Van der Giessen, 2015; Cui et al., 2020, 2016; Fan et al., 2021; Zhou et al., 2010). To specifically examine the effect of the applied strain rate, additional simulations were conducted at a strain rate of $10^2$ $s^{-1}$ with all other settings unchanged. The lack of statistically significant differences observed in these simulations demonstrates that the applied strain rate does not influence the subsequent discussion or the derived conclusions, particularly concerning the relative strain burst magnitude and strain hardening trends observed between low and high dislocation densities.

**Table 1.** Material parameters of MgO.

| Type | Symbol | Meaning | Value | Unit | Ref. |
|---|---|---|---|---|---|
| Material | $\mu$ | Shear modulus | 133 | GPa | - |
| | $\nu$ | Poisson's ratio | 0.18 | - | - |
| Dislocation mobility | $\tau_p$ | Peierls stress | 150 | MPa | (Amodeo et al., 2011) |
| | $\Delta H_0$ | Critical enthalpy for kink pair nucleation | 1.14 | eV | (Amodeo et al., 2011) |
| | $p$ | Fitting parameters | 0.5 | - | (Amodeo et al., 2011) |
| | $q$ | Fitting parameters | 2 | - | (Amodeo et al., 2011) |
| | $l_c$ | Critical width of kink pairs | 113 | $b$ | (Amodeo et al., 2011) |
| | $B_e$ | Drag coefficient for edge dislocation | $1\times10^{-4}$ | Pa·s | (Reali et al., 2017) |
| | $B_s$ | Drag coefficient for screw dislocation | $5\times10^{-4}$ | Pa·s | (Reali et al., 2017) |
| | $T_a$ | Athermal transition temperature | 600 | K | (Amodeo et al., 2011) |

## 3. Results and Analyses

### 3.1. Mechanically seeded dislocations with different densities

Detailed analysis of dislocation types and structures within the wear tracks is given in Figure 1 across varying scratching passes. The dislocation density of the reference sample (unscratched sample) was ~7 × $10^{11}$ $m^{-2}$ (~$10^{12}$ $m^{-2}$, possibly introduced during crystal growth or sample polishing), estimated by the chemically etch pits method, as illustrated in our previous study (Preuß et al., 2024). The dislocation distribution of the reference sample was revealed in Figure 1a, whose dislocation density is higher than $SrTiO_3$ (Zhang et al., 2025). After 1× scratching, the dislocation density in (001) MgO increases significantly to ~$10^{14}$ $m^{-2}$ on average, as illustrated in Figure 1b and our previous studies, demonstrating a clear gradient distribution. At a depth of ~500 nm beneath the surface, the dislocation density is much higher with ~$10^{15}$ $m^{-2}$. Beyond 1.5 µm, dislocations predominantly exhibit 45° line structures, indicative of activation on {110} planes (Figure 1d). The dislocations were penetrated to a depth of ~150 µm, as displayed in our previous study (Preuß et al., 2024). This gradient distribution highlights the inward progression of dislocation motion and multiplication during the scratching process.

With 10× scratching passes, the average dislocation density surpasses ~$10^{15}$ $m^{-2}$, as demonstrated in Figure 1c. At a depth of ~2 µm, cell-like dislocation structures size of 200-300 nm emerge (Figure 1e), signifying that the dislocation cell structure has expanded significantly with the increased scratching passes, while such dislocation cell structures were not detected in $SrTiO_3$ after 20× scratching passes for their lower dislocation density (~$10^{14}$ $m^{-2}$) (Zhang et al., 2025). Due to the high density of dislocation cell structures, it is difficult to calculate their density by line interaction methods accurately, so here we only give an order of magnitude. Following this extensive scratching, the dislocations primarily form cellular structures rather than 45° dislocation lines due to enhanced interactions during the bi-direction scratching. Despite the high dislocation density and structural changes, the surface retains its (001) single crystal structure without forming (sub)grains, as depicted in Figure S4 in the Supplementary Materials. This is much different from the grain refinement caused by scratching in metals (Greiner et al., 2016). Figure 1f further summarizes the relationship between dislocation density and scratching pass numbers, demonstrating a continuous increase in dislocation densities. Note that the dislocation densities in Figure 1 were estimated at a depth of approximately 5 to 6 µm, covering the micro-pillars' height. For the scratched sample, the gradient-distributed dislocation density was assessed through different depths, as revealed in our previous (Preuß et al., 2024). These results show the effectiveness of scratching in tuning dislocation densities and structures while preserving the overall single-crystal integrity of the MgO surface.

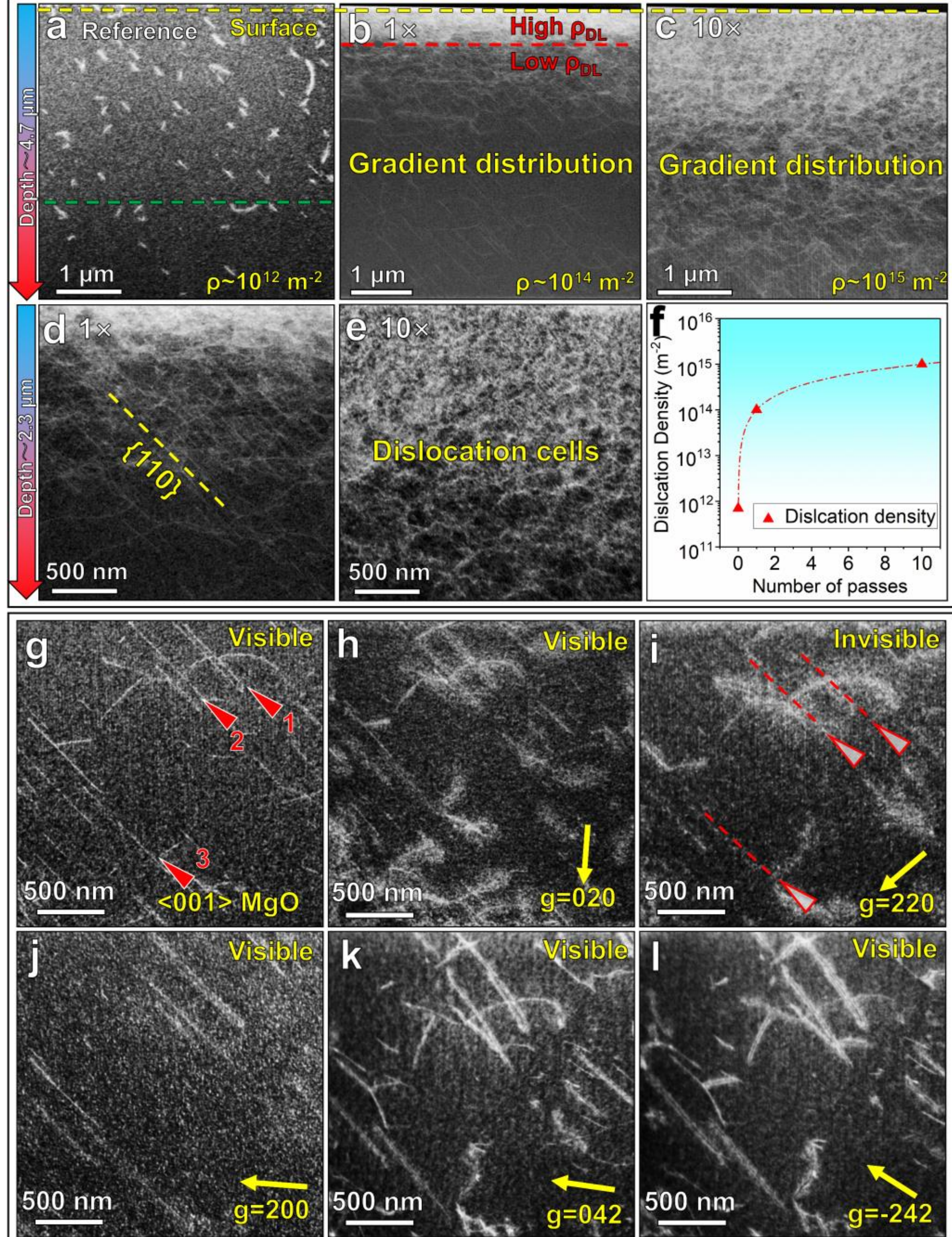

**Figure 1.** Dislocation structure and distribution tuned by varying scratching pass numbers. (a) Dislocation distribution in the reference sample (0× scratching); (b-c) gradient-distributed dislocation structures after 1× and 10× scratching, respectively; (d-e) zoomed-in images showing higher dislocation density near the surface of the 1× scratched sample and the formation of dislocation cells structures in the 10× scratched sample; (f) variation in dislocation densities as a function of the number of scratching passes, estimated using the TEM images. The arrow on the left indicates the direction of the height of the micro-pillar from top to bottom, and the green dashed line in (a) indicates the depth and the bottom of the micro-pillar. Experimental tilting series

ADF-STEM images under different conditions indicate the edge-type dislocations induced by 1× scratching. (g) Dislocation structure observed along the <001> zone axis, labeled by the blue rectangle; (h) dislocation visible with g=020; (i) dislocation invisible with g=220; (j) dislocation visible with g=200; (k) dislocation visible with g=042 and (l) g=$\bar{2}42$. The parameter $\rho_{DL}$ represents dislocation density.

Dislocation invisibility analysis and trace analysis were utilized to confirm the Burgers vector and line vectors of the scratching-induced dislocation structures in the (001) MgO single crystal. Dislocations become invisible when the Burgers vector **b** is perpendicular to the diffraction vector **g** (**g** × **b**=0). The dislocation line vector **t** was determined by the trace analysis across different non-coplanar zone axes. Figures 1g-1l illustrates the contrast variation and morphology of 1× scratching-induced dislocation lines under different diffraction vectors **g**. As marked by the red triangles in Figure 1g-l, dislocations labeled 1-3 are visible along the <001> zone axis and show straight 45° lines. These dislocations 1-3 became invisible when the **g** vector is 220 but remain visible in other cases in Figure 1g-l (1h, 1j, 1k, and 1l). Dislocation invisibility analysis (Table S1) reveals that the Burgers vector for straight lines dislocations 1-3 is ± a[$1\bar{1}0$]. Trace analysis along the <001> axis and <0$\bar{1}$2> axes yields the line vector **t** = ± a[$1\bar{1}0$]. Since Burgers vector **b** is parallel to line vector **t**, dislocation lines 1-3 are identified as screw-type dislocations.

### 3.2. Micro- and nano-pillar compression

#### 3.2.1 Impact of mechanically seeded dislocations with different densities

Room-temperature micro-pillar compression tests were performed to investigate the effects of varying dislocation densities on the micro-mechanical properties of (001) MgO single crystal, as demonstrated in Figure 2 and Figure 3. Initially, micro-pillars with increased dislocation densities were compressed to ~25% total strain to assess their plastic deformation behavior (Figure 2). A separate set of samples was then compressed to 25% (reference sample) and 40% (1× and 10× scratched samples) to compare the fracture modes resulting from increased dislocation densities, as displayed in Figure 3. In the reference sample with a dislocation density of approximately $\sim 10^{12}$ $m^{-2}$, micro-pillar 1 displayed post-mortem plastic deformation features with 22% compression strain (Figures 2b and 2c). Several coarse slip bands appeared on the pillar surface, with noticeable multi-slip along the (011) and (0$\bar{1}$1) planes at the base (Figure 2b). The corresponding stress-strain curve revealed almost constant stress after the yielding (at ~1 GPa) but with large load drops (Figure 2d). In contrast, micro-pillar 2 from the same reference sample developed substantial crack formation along the (011) plane after 22% deformation, resulting in fracture (see later Figure 3b). This simultaneous occurrence of plastic deformation and fracture is attributed to the uneven distribution of dislocations, as noted in Figure 1a. Given the small size of these micro-pillars (1 μm in diameter and 3 μm in height), they contained only a few dislocations, resulting in insufficient plasticity and occurrence of cracks. In comparison, micro-pillars with higher dislocation densities exhibited no cracking, even after 25% deformation.

After 1× scratching, the (001) MgO single crystals developed a gradient-distributed dislocation structure with an averaged density of ~ $10^{14}$ $m^{-2}$ over a depth of ~3 μm (Figure 1b). These micro-pillars, with mechanically seeded gradient-distributed dislocations, underwent 25% deformation without cracking. Micro-pillar 1 from the 1× scratched sample deformed plastically, forming a high density of slip bands along the (011) surface (Figure 2f), which appeared smoother compared to the coarse slip bands of the reference micro-pillar 1 (Figure 2b). Thicker slip traces indicate more homogenous deformation in the 1× scratched micro-pillar.

With 10× scratching, the (001) MgO micro-pillars developed dislocation cellular structures with a dislocation density of ~ $10^{15}$ $m^{-2}$ near the surface (Figure 1c). After 25% deformation, the micro-pillar subjected to 10× scratching showed plastic deformation without cracking (Figures 2j and 2k). The yield strength is ~1.60 GPa. The slip bands were finer and less pronounced compared to those observed in the 1× scratched crystals (Figure 2f). As dislocation density increased, the plastic flow transitioned from the intermittent feature (Figures 2d and 2h) to continuous hardening (Figure 2l). The absence of significant stress drops correlated with the finer slip traces observed after compression suggests that higher dislocation densities improve plastic deformation stability and reduce localized stochastic deformation events.

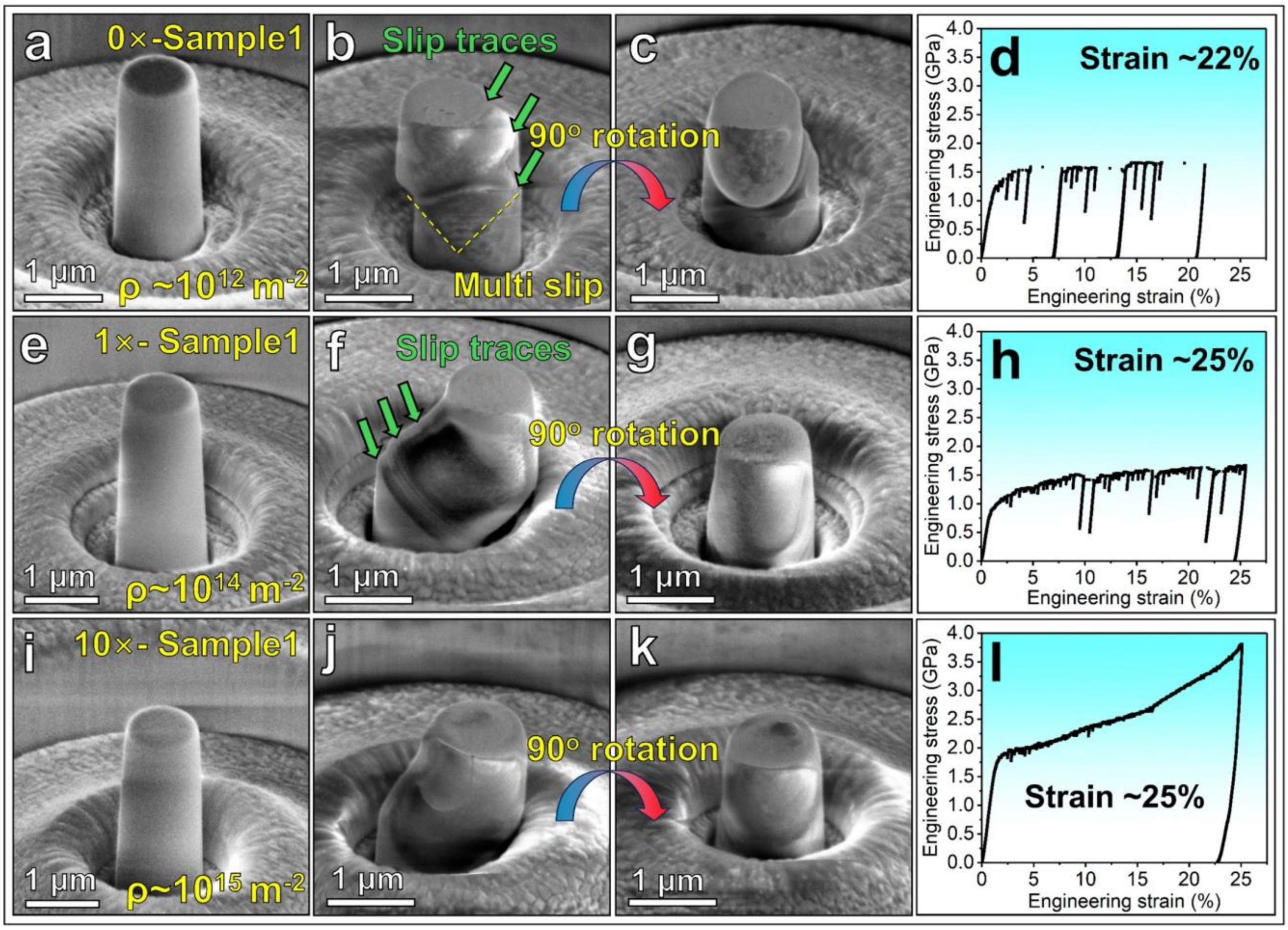

**Figure 2.** Representative *ex situ* room-temperature micro-pillar compression results of the (001) MgO single crystals with a diameter of 1 μm and varying mechanically seeded dislocation densities without electron beam effect, all without fracture. (a) SEM images of the reference micro-pillar (sample 1: 0× scratching) before compression; (b) and (c) deformed reference micro-pillar showing multi-slip and high-density slip traces; (d) engineering stress-strain curve of the plastically deformed reference micro-pillar; (e) SEM images of the 1× scratched micro-pillar before compression; (f) and (g) deformed 1× scratched micro-pillar displaying high-density slip traces; (h) engineering stress-strain curve of the 1× scratched micro-pillar; (i) SEM images of the 10× scratched micro-pillar before compression; (j) and (k) deformed 10× scratched micro-pillar showing minimal slip traces; (l) engineering stress-strain curve of the 10× scratched micro-pillar.

The micro-pillar with a mechanical seeded dislocation density of approximately ~$10^{14}$ $m^{-2}$ (1× scratching) experienced deformation up to 40%, as displayed in Figures 3e-h. The yield strength is about 0.88 GPa. Following this deformation, the micro-pillar exhibited a higher density of slip traces (Figure 3f), accompanied by the initiation and propagation of nano-cracks along the surface edges (Figure 3f). Deformation was dominated by shear and bending along the (011) plane, indicative of a single slip mechanism. As the micro-pillar bent to the right, stress concentration developed in the middle, causing nano-cracks to propagate along the (011) plane in Figure 3f. The bending angle reached ~26°, and the micro-pillar came into contact with the matrix after ~34% deformation. This resulted in a dramatic increase in compressive stress (Figure 3h).

In contrast, the micro-pillar with a dislocation density of ~$10^{15}$ $m^{-2}$ (10× scratching) exhibited a different fracture behavior after 40% strain. The yield strength reaches ~1.40 GPa. Nano-cracks propagated along the <010> direction at the top surface, as demonstrated in Figure 3j. Compared with the 1× scratched micro-pillar, the 10× scratched micro-pillar displayed finer slip traces and a reduced bending angle, reaching only ~9° at 40% strain—a reduction of about 65% compared to the 1× scratched micro-pillar (Figure 3f). The continuously compression caused the diameter at the top of the deformed pillar to increase to approximately 1.3 μm (Figure 3j), contributing to a rise in compressive stress (Figure 3l). Similar to the 1× scratched micro-pillar (40% strain), contact with the matrix occurred after ~34% strain, leading to a sharp increase in stress (Figure 3l). These results highlight the influence of dislocation density on deformation behavior, with higher dislocation densities (10× scratched) resulting in finer slip traces, reduced bending angles, increased resistance to deformation, and a higher yield strength compared to lower dislocation densities (1× scratched).

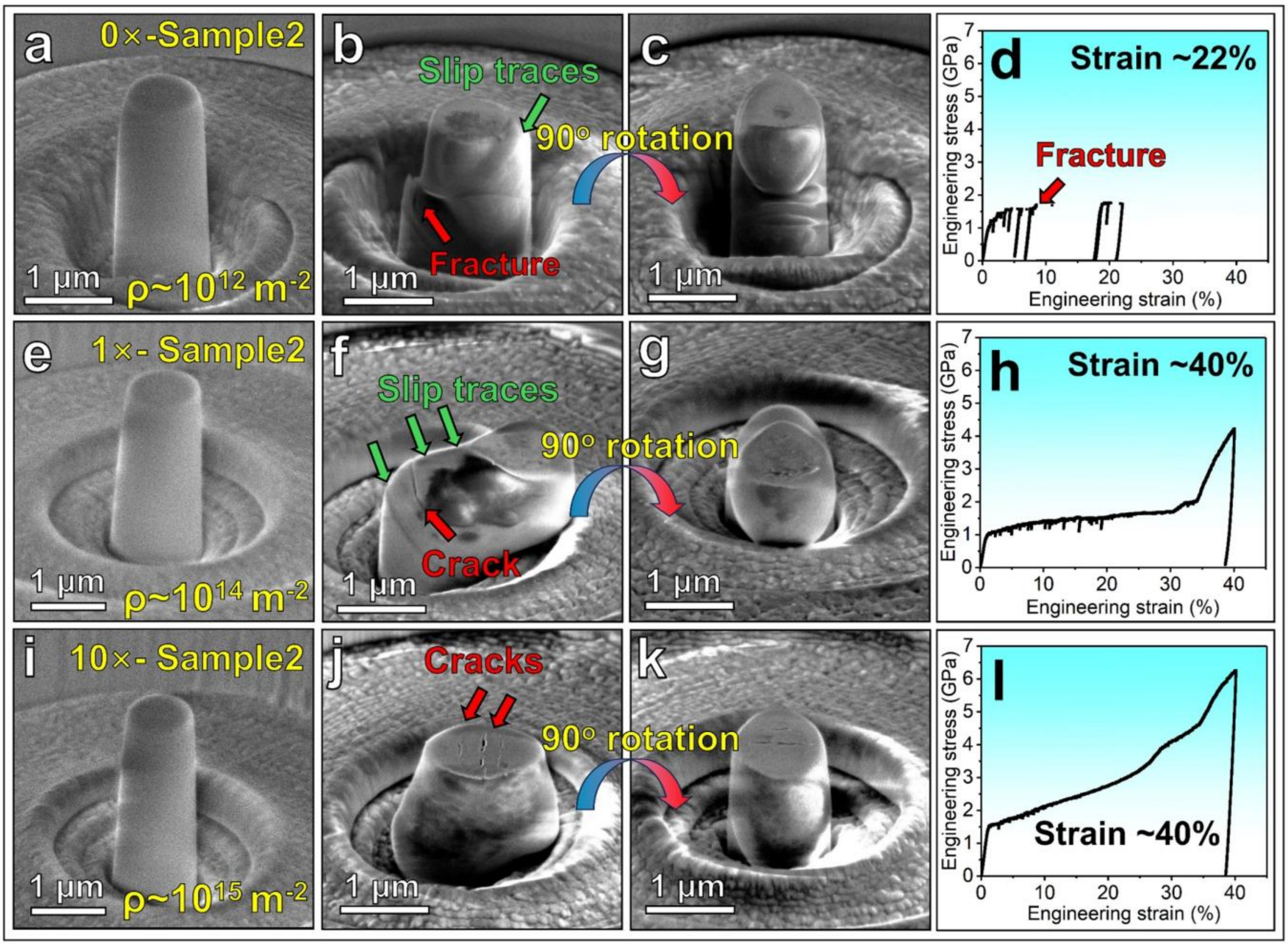

**Figure 3.** Representative *ex situ* room temperature micro-pillar compression results of the MgO single crystals exhibit cracks without electron beam effect. (a) SEM images of the reference micro-pillar (sample 2: 0× scratching) before compression; (b) and (c) deformed reference micro-pillar showing fracture along the (011) plane; (d) engineering stress-strain curve of the fractured reference micro-pillar; (e) SEM images of the 1× scratched micro-pillar before compression; (f) and (g) deformed 1× scratched micro-pillar show high-density slip traces and cracks on the edge of the surface; (h) engineering stress-strain curve of the 1× scratched micro-pillar 2; (i) SEM images of the 10× scratched micro-pillar before compression; (j) and (k) deformed 10× scratched micro-pillar show unnoticeable slip traces and lateral cracks on the middle of the top surface; (l) engineering stress-strain curve of the 10× scratched micro-pillar 2.

### 3.2.2. Size effect with varying dislocation densities

To investigate the size effect on mechanical properties, micro-pillar compression tests were performed on (001) MgO single crystals with varying dislocation densities and diameters, both inside and outside the scratched tracks, as demonstrated in Figure 4. Similar to ductile metals, Figures 4a and 4b revealed a pronounced size effect. Outside the scratched track (reference region), micro-pillars with a diameter of 1 μm exhibited strong stress drop and intermittent plastic flow behavior in their engineering stress-strain curves (Figure 4a). As the diameter increased to 5 μm, the strain bursts became negligible, and the stress drops became finer, indicating a more stable deformation mechanism. Additionally, the yield strength decreased significantly from ~1 GPa for the 1 μm micro-pillar to ~0.4 GPa for the 5 μm micro-pillar. Cracks in the 5 μm micro-pillar propagated after 10% strain (Figure 4d), while in the 1 μm micro-pillar, the crack formation occurred at ~22% strain (Figure 3b). After ~20% strain, the 5 μm-diameter micro-pillar developed several cracks at the top of the pillar (Figure 4e).

For micro-pillars with mechanically seeded dislocation densities of ~$10^{15}$ $m^{-2}$, the average yield strength of the 1 μm micro-pillar was approximately 1.50 ± 0.07 GPa. Note this sample (10×, 1 μm-3) deviated slightly from the scratch center compared to the 1 μm 1-2 samples, as shown in Figure S11 below, resulting in a slightly lower dislocation density and consequently a slightly lower yield strength but still within the margin of error. However, as the diameter increased to 5 μm, the yield strength decreased to ~0.6 GPa upon yielding, and the 5 μm micro-pillar exhibited multi-slip behavior along the {110} planes (Figures 4g-h). After 20% strain, cracks propagated in an array along the [100] direction (Figure 4c), consistent with the crack propagation behavior shown in Figure 3j. It should be mentioned that the 5 μm micro-pillars did not exhibit significant different in mechanical behavior under electron beam or out of electron beam. The reference 5 μm-2 (ρ ~$10^{12}$ $m^{-2}$, red curve in Figure 4a deformed to ~20%) and the 10× scratched 5 μm-3 (ρ ~$10^{15}$ $m^{-2}$, red curve in Figure 4b) were both tested outside the SEM without the influence of the electron beam. The other 5 μm micro-pillars in Figures 4a and 4b were tested under SEM, their stress-strain curves did not show significant difference with and without beam. Compared with the 5 μm micro-pillars with a lower pre-existing dislocation density of ~$10^{12}$ $m^{-2}$, those with higher dislocation densities displayed much finer slip traces in Figures 4g and 4h, which correlated with the finer stress drop behavior observed in Figure 4b. These observations highlight the impact of both size and dislocation density on the mechanical behavior of MgO micro-pillars. *In situ* micro-pillar compression tests on (001) MgO micro-pillars with a diameter of 5 μm are demonstrated in the Supplementary Videos 1-2. A detailed discussion of the size effect will follow in **Section 4**.

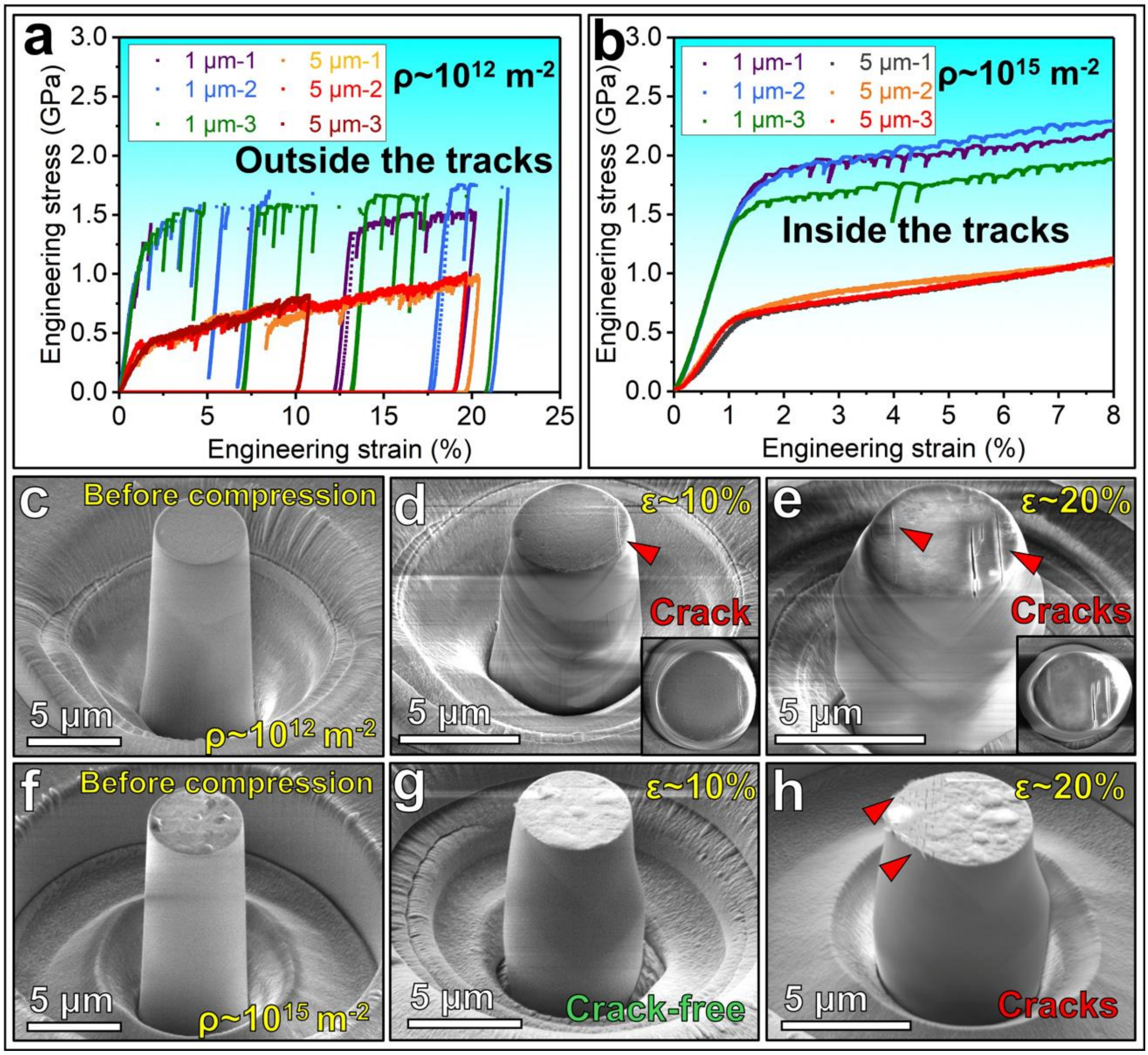

**Figure 4.** Size effect in the (001) MgO micro-pillar compression tests inside and outside the scratched tracks, the 5 μm diameter pillars were tested inside the SEM. (a) Engineering stress-strain curves of the (001) MgO micro-pillars outside the scratched tracks with diameters of 1 μm and 5 μm; (b) engineering stress-strain curves of the (001) MgO micro-pillars inside the scratched tracks with diameters of 1 μm and 5 μm; (c) SEM image of the micropillar outside the scratched track before compression; (d) MgO micro-pillar with pre-existing dislocation density of ~$10^{12}$ $m^{-2}$ deformed to ~10%, with the inset revealing a crack propagating at the top; (e) micro-pillar with pre-existing dislocation density of ~$10^{12}$ $m^{-2}$ deformed to ~20%, with the inset showing an array of cracks at the top; (f) SEM image of the micro-pillar inside the scratched track with a mechanical seeded dislocation density of ~$10^{15}$ $m^{-2}$ before compression; (g) micro-pillar inside the scratched track with a dislocation density of ~$10^{15}$ $m^{-2}$ deformed to 10%; and (h) micro-pillar deformed to ~20%, cracks propagated at the top.

We then performed nanoscale *in situ* pillar compression tests under TEM in order to reveal the dynamic behavior of mechanically seeded dislocations on the plastic deformation of MgO, also to see the size effect with smaller diameter, as illustrated in Figure 5 and Supplementary Videos 3-4. All the nano-pillar are compressed along <001> axis. Figures 5a-b demonstrate the *in situ* compression result of the ~400 nm diameter pillar with mechanically seeded dislocation density of ~$10^{15}$ $m^{-2}$. Due to the high density of dislocation cell structures (Figure 1c), the dislocations in Figures 5a and 5c show indistinguishable contrast before deformation. The ~400 nm pillar deformed elastically until ~5% strain, as shown in Figures $5a_2$ and 5b. Slip occurred as the compression strain increased to ~12% and ~15%, as indicated by the green arrows in Figures $5a_3$ and $5a_4$. The ~400 nm pillar deformed to ~25% without cracking, as displayed in Figure $5a_4$. The ~400 nm pillar finally deformed to ~70% (Figure S6 and Supplementary Video 3), and the yield strength reached ~2.35 GPa. Nevertheless, the pillar ~400 nm with none-seeded dislocation fractured after ~13% total strain (Figure S7). For the nano-pillar with a diameter of ~500 nm, slip traces were detected with increased compression strain up to ~9.2% (Figure $5c_3$), ~15.3% (Figure $5c_4$), and ~21% (Figure $5c_5$) without cracking. The yield strength of the ~500 nm pillar decreased to ~1.57 GPa.

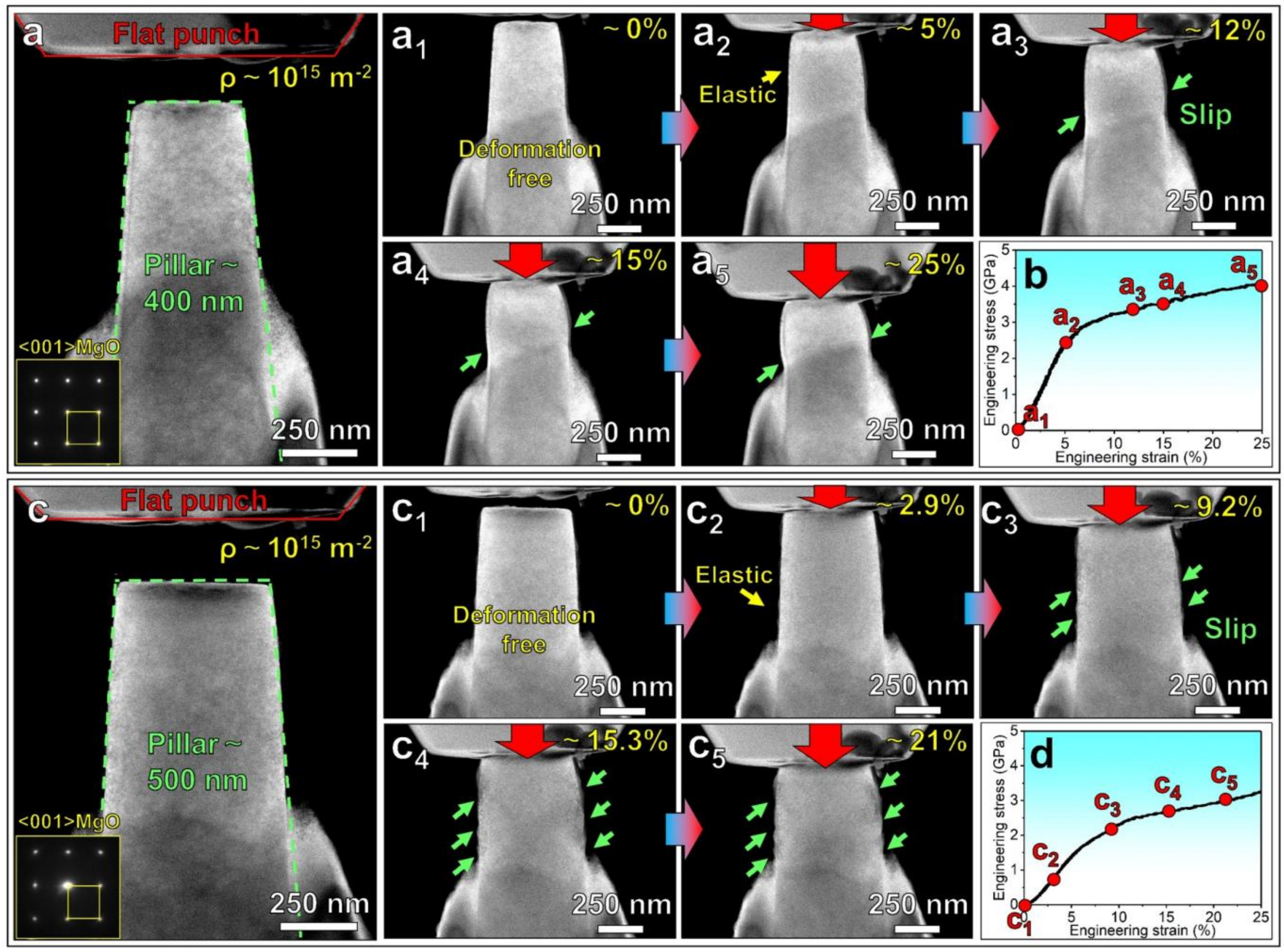

**Figure 5.** TEM *in situ* nano-pillar compression test results (10× scratched, ρ ~$10^{15}$ $m^{-2}$) with diameters of ~400 nm and ~500 nm. (a) Dislocation distribution of the <001> MgO pillar with a diameter of ~400 nm before deformation; ($a_{1\text{-}5}$) TEM snapshots of the ~400 nm pillar with increased plastic strain up to ~25%; (b) engineering stress-strain curve of the ~400 nm pillar; (c) dislocation distribution of the <001> MgO pillar with a diameter of ~500 nm before deformation; ($c_{1\text{-}5}$) TEM snapshots of the ~500 nm pillar with increased plastic strain up to ~21%; (d) engineering stress-strain curve of the ~500 nm pillar.

The mechanical properties of the MgO micro-pillars with varying mechanically seeded dislocation densities and diameters are summarized in Table 2. The yield strengths of the MgO micro-pillar with 1 μm diameter initially decreased to ~0.8 GPa at a dislocation density of ~ $10^{14}$ $m^{-2}$ before rising to ~1.5 GPa at a density of ~$10^{15}$ $m^{-2}$. These findings emphasize the combined influence of size and dislocation density on the mechanical properties and deformation behavior of MgO single crystals.

**Table 2** Mechanical properties of the (001) MgO micro-pillars with different dislocation densities.

| Scratching passes | Diameter (D, μm) | Dislocation density ($m^{-2}$) | Yield strength ($\sigma_{0.2}$, GPa) | Predicted yield strength ($\sigma_{0.2}$, GPa) | Compression strain (%) |
|---|---|---|---|---|---|
| 0× | 1 | ~$10^{12}$ | 0.98±0.04 | / | 22% |
| 0× | 5 | ~$10^{12}$ | 0.37±0.03 | / | 20% |
| 1× | 1 | ~$10^{14}$ | 0.82±0.06 | 0.82 | 40% |
| 10× | 0.4 | ~$10^{15}$ | 2.35 | / | ~70% |
| 10× | 0.5 | ~$10^{15}$ | 1.57 | / | ~30% |
| 10× | 1 | ~$10^{15}$ | 1.49±0.06 | 1.50 | 40% |
| 10× | 3 | ~$10^{15}$ | 0.50±0.02 | / | 15% |
| 10× | 5 | ~$10^{15}$ | 0.58±0.02 | / | 20% |

### 3.2.3. Cross-sectional view of dislocation structure change

In the low dislocation density region (Figure 3b), the micro-pillar exhibited shear deformation along the (011) plane, resulting in crack formation and subsequent fracture. As the dislocation density increased, cracks propagated along the sides of the surface after 40% strain in the 1× scratched sample, primarily due to bending and stress concentration. At a higher dislocation density of ~$10^{15}$ $m^{-2}$, cracks propagated along the <010> direction after deformation.

TEM specimens were prepared from the micro-pillar subjected to 10× scratching and 40% deformation, with the TEM lamellae lifted from the middle of the micro-pillar along the scratching direction to observe crack propagation directly. The TEM results (Figure 6) revealed a significant increase in dislocation density after 40% strain compared to the as-scratched state (Figure 1c). However, accurately quantifying the dislocation density was challenging due to the diminished contrast between dislocation lines and the surrounding matrix at such high dislocation densities. Following deformation, dislocations accumulated on the (011) and $(0\bar{1}1)$ planes, aligning with the high-density slip traces observed. As shown in Figure 6b, surface nano-cracks propagated along the (010) plane at the top of the micro-pillar (Figure 3j). After 40% deformation, the dislocation structure transitioned from cellular arrangements to highly entangled dislocation lines on the (011) and (0$\bar{1}$1) planes. This entanglement and pile-up of dislocation lines across two intersecting slip planes are believed to be the primary contributors to crack initiation (Figure 6c), a mechanism also reported in SiC micro-pillar compression tests (Amodeo et al., 2018). The green dashed lines in Figures 6c and 6d indicate the Moiré patterns, which are related to the generation of local strain near the crack.

Beneath the crack tip, lattice torsions were identified (Figures 6d and 6e). Additionally, several intersecting moiré fringes appeared beneath the crack tip, highlighted by green dashed lines in Figure 6d. The presence of dislocation dipoles post-deformation indicated a Burgers vector of 1/2[110], corresponding to the activation of the 1/2<110>{110} slip system (Figure 6f). These findings highlight the role of dislocation entanglement, slip system activation, and lattice distortions in driving crack propagation in high dislocation density regions.

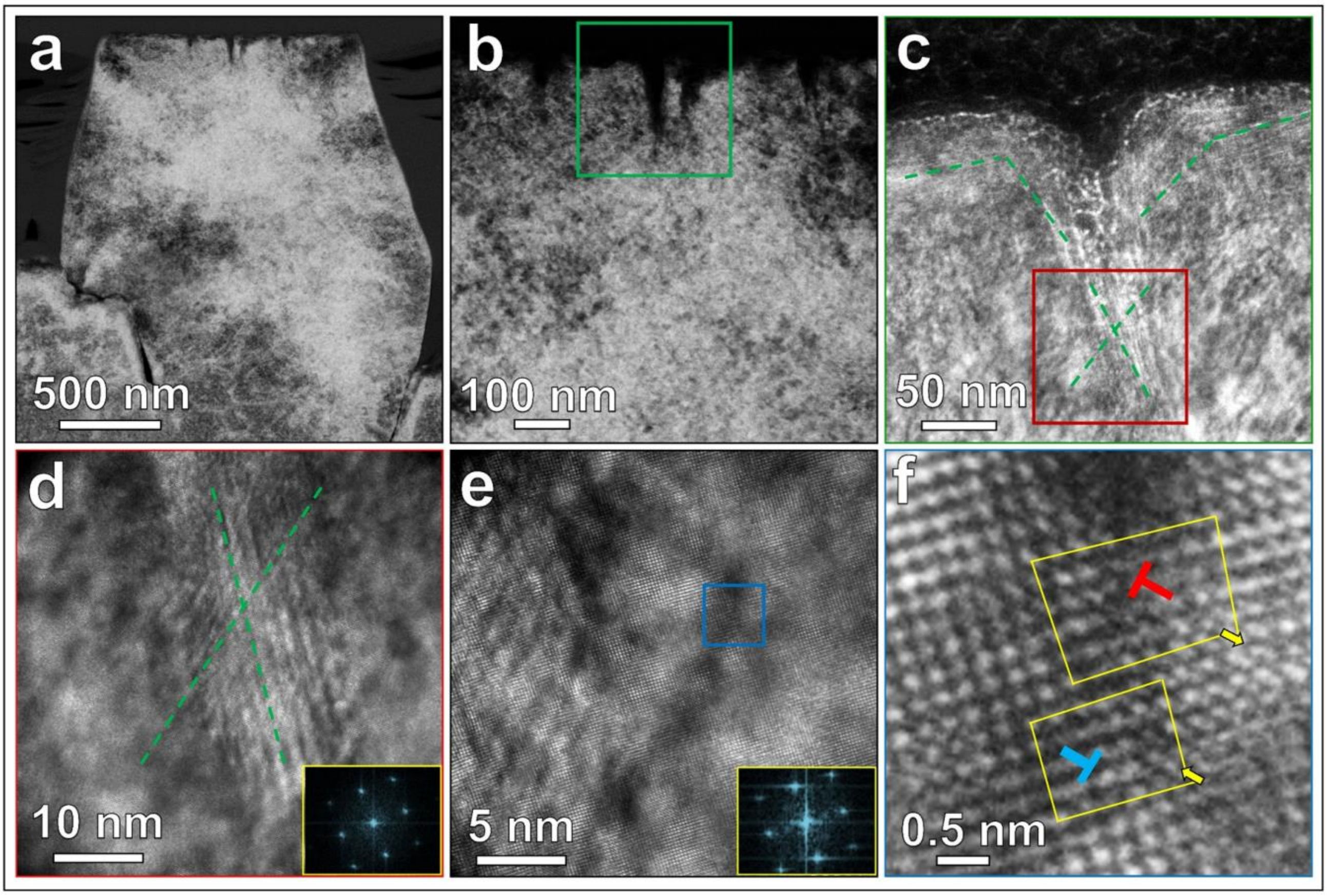

**Figure 6.** ADF micrographs and HRTEM analysis of the micro-pillar after 40% deformation shown in **Figure 3j**. The TEM lamella was lifted along the scratching direction. (a) Cross-sectional TEM image of the entire

deformed micro-pillar; (b) high-density dislocations and surface crack penetration observed at the top of the micro-pillar post-deformation; (c) magnified TEM image of the region shown in (b); (d) a HRTEM image revealing the crack tip structure; (e) orientation of the side of the crack tip; (f) a HRTEM showing the edges dislocations structure in the region marked in blue box in (e).

Figure 7 illustrates the atomic imaging of the MgO dislocation core structures of the deformed micro-pillar in Figure 6. A high-resolution HAADF-STEM image and simulated image of a dislocation-free MgO sample confirm a perfect cubic structure without any dislocations (Figures 7a and 7b). The simulated HRTEM image of the predicted MgO structure was processed by using JEMS software for a sample thickness of 20 nm (Lay, 2013). The fast Fourier transform (FFT) pattern in Figure 7c reveals the orientation of the (001) MgO. As displayed in Figures 7d and 7g, edge dislocations with Burgers vector of ±[110] are detected in the deformed MgO sample. The corresponding inverse fast Fourier transform (IFFT) pattern in Figures 7e and 7h highlights the misaligned atomic planes in Figures 7d and 7g, respectively. The geometric phase analysis (GPA) in Figures 7f and 7i visualize the strain field around the dislocation core.

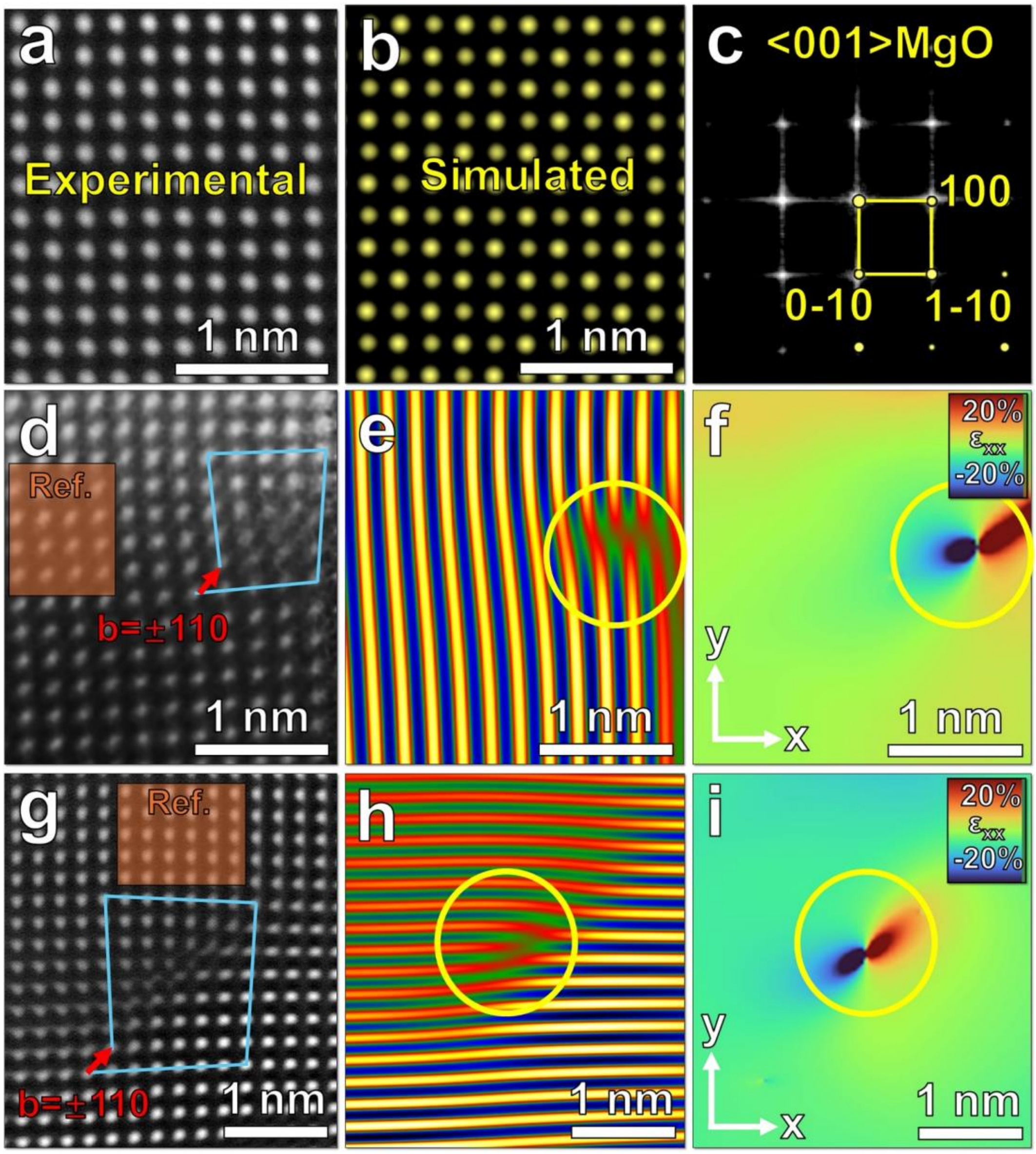

**Figure 7.** Atomic resolution imaging of the deformed MgO sample. (a) Experimental HAADF image of the dislocation-free MgO matrix; (b) Simulation of atomic arrangements in the dislocation-free MgO by using JEMS software (Lay, 2013); (c) FFT pattern in (a) reveal the <001> orientation of the MgO sample; (d) HAADF-STEM image illustrates the edge-type dislocations with Burgers vectors of ±[110]; (e) IFFT pattern in (d); (f) GPA reveals the strain field around the dislocation core in (d); (g) HAADF-STEM image illustrates the edge-type dislocations with Burgers vectors of ±[110]; (h) FFT pattern in (g); (i) GPA depicts the strain field around the dislocation core in (g).

### 3.3. Bulk compression

Since brittle materials exhibit significant heterogeneity at different scales due to their complex microdefects, and their macroscopic mechanical behavior strongly depends on crack initiation, propagation, and friction slip (Kong et al., 2023). Transitioning from micro-pillar to bulk compression in MgO single crystal (dimension of 3 mm in length and width, 6 mm in height) bridges the gap between microscale and macroscale dislocation-driven plastic deformation behaviors, as illustrated in Figure 8. The *in situ* deformation process is displayed in Supplementary Video 5. While micro-pillar compression tests focus on individual dislocation motion and dislocation-density dependent mechanical behavior, bulk compression tests reflect more realistic defect interactions and offer insights into size effects. Here, we use the average dislocation spacing ($L = 1/\sqrt{\rho}$) instead of dislocation density to compare with the sample's characteristic length $D$ (the value is 3 mm in this study), since dislocation density is a volume-scaling invariant parameter (Fang et al., 2024). The engineering stress-strain curve shown in Figure 8a reveals a yield strength of ~126 MPa with an average dislocation spacing $L$ of approximately 1–3 μm, highlighting a significant size effect when compared with the micro-pillar compression tests. Initially, as depicted in Figure 8b, the sample exhibits no visible microcracks or slip bands on its surface or internally. However, as the compression strain reaches around 2.9%, numerous slip bands appear within the sample (Figure 8c), marked by the green and blue arrows that represent different slip orientations. The green arrows highlight slip bands originating from the <010> direction, lying along the (101) and $(\bar{1}01)$ planes and intersecting one another. The blue arrows indicate the presence of several 45° slip traces generated from the (011) and $(0\bar{1}1)$ planes, reflecting a multi-slip mechanism. These observations align with the activation of the 1/2<110>{110} slip systems, which are characteristic of MgO under compression.

At a strain of approximately 3.6%, a significant macroscopic crack forms in the center of the sample, denoted by the red triangle in Figure 8d. This primary crack propagates along the (001) plane, located at the intersection of two sets of intersecting 45° slip bands. This suggests that interactions between different slip planes contribute to crack initiation, corresponding to the first stress drop observed in the engineering stress-strain curve (Figure 8a). As the strain increases to ~4.6%, macroscopic cracks expand, and secondary cracks begin to develop along the <001> direction (Figure 8e). At around 7.3%, the density of slip bands along the <010> direction continues to increase, leading to further crack formation (Figure 8f). Eventually, the sample undergoes complete fracture after ~11% deformation, as demonstrated in Figure 8g.

Surface morphology imaging provides critical insights into linking the macroscopic mechanical response of (001) MgO single crystals with their microscopic deformation mechanisms, facilitating a deeper understanding of their plastic behavior. Figures 8h-k presents the optical microscopy images of a (001) MgO bulk sample deformed to a strain of 2%. The side surfaces of the deformed sample display numerous 45° slip traces on the (011) and $(0\bar{1}1)$ planes, as shown in Figures 8h and 8i. These slip lines, indicated by the blue arrows in Figure 8i, exhibit a slight angular shift rather than forming perfectly straight lines. Moreover, the slip bands along the (011) and $(0\bar{1}1)$ planes are composed of multiple short segments, each approximately 30–50 μm in length, rather than continuous lines. On the front face of the deformed sample (Figure 8k), high-density slip traces parallel to one another are observed, which aligns with the findings in Figures 8c and 8d. These parallel slip bands emerge where the 45° slip bands from the sides of the samples terminate on the front face of the sample. Furthermore, the top of the sample (Figure 8j) shows mutually perpendicular slip bands along the <001> and <010> directions, further reinforcing the presence of intersecting 45° slip bands. These observations underline the complex interaction of slip systems during deformation, where intersecting and parallel slip bands link the macroscopic strain response to the underlying microscopic plastic deformation mechanisms. This provides valuable insights into the anisotropic deformation behavior of (001) MgO under compressive loading.

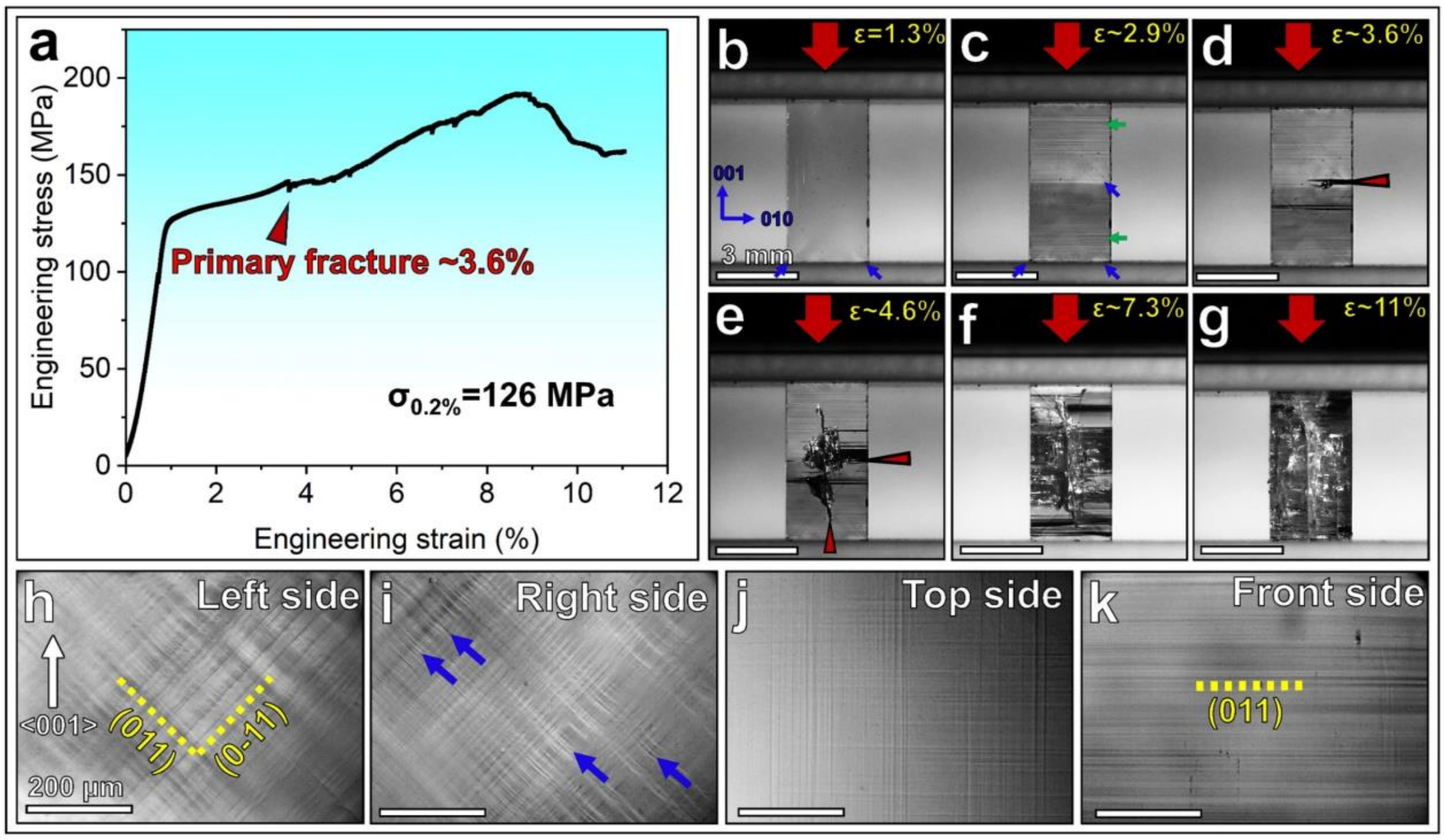

**Figure 8.** Room-temperature bulk compression tests of the (001) MgO single crystal with an average dislocation

spacing $L$ of approximately 1–3 μm under a strain rate of $1.5\times10^{-4}$ $s^{-1}$. (a) Engineering stress-strain curve; (b-g) *in situ* bulk compression images at various strain levels. The scale bar in (b) applies to figures c-g. The green and blue arrows represent the directions of the slip traces, while red triangles denote the initiation and propagation of cracks. (h-i) OM images of the left and right sides of the deformed sample, displaying high-density 45° cross-hatched slip traces on the {110} planes. Blue arrows indicate the shifts of the slip traces. (j-k) Top and front views of the sample show parallel slip traces formed during deformation. The scale bar in (h) applies to figures i-k.

DIC was employed to analyze the strain distribution and identify localized deformation during bulk compression, providing insights into the deformation process. DIC offers significant advantages compared to traditional compression tests, including full-field, real-time strain distribution and deformation mapping. This capability is particularly valuable for studying dislocation-induced plasticity in ceramic materials, as it enables precise identification of localized deformation zones, strain heterogeneity, and slip band formation (Guo et al., 2023). By visualizing strain evolution, DIC reveals how dislocation motion influences overall plasticity through macroscopic strain field analysis, offering detailed insights into the mechanisms governing deformation.

Figure 9 illustrates 2D strain distribution maps measured using DIC during the compression at a strain rate of $1.5\times10^{-4}$ $s^{-1}$. The distribution of color maps usually does not fully cover the entire sample due to the balance of resolution when calculating the strain field. At an initial strain of 2.5% strain (Figure 9a), the sample exhibits minimal stress concentration, indicating uniform deformation. As the strain increases to 5% (Figure 9b), localized strain intensifies at the center of the sample and the top of the sample, signaling potential crack initiation, similar to the behavior observed in Figure 8e. At strains of 7.5% and 8% (Figures 9c and 9d), a distinct macroscopic crack becomes visible, marked by a red dashed line and propagating along the <001> direction. With further increases in strain to 8.2% and 9.0% (Figures 9e and 9f), crack propagation intensifies, while strain remains primarily in the central region of the sample. As the deformation progresses to 11% and 14% (Figures 9g and 9h), the strain begins to propagate outward from the center toward the edges of the sample, reflecting a broader distribution of deformation as cracks grow and spread. The dynamic evolution of strain and crack formation during compression is further illustrated in Supplementary Video 6. This analysis highlights the effectiveness of DIC in capturing localized deformation and tracking crack propagation, providing critical insights into the dislocation-mediated plasticity and failure mechanisms of (001) MgO under bulk compression.

To reveal the failure mechanisms, dislocation behavior, and their role in deformation, post-deformation analyses using optical microscopy (OM) and TEM were demonstrated in Figures 9i-k. The OM images reveal the expansion of microscopic cracks that were not visible during the lower magnification *in situ* macroscopic compression tests in Figure 8. TEM characterization, on the other hand, highlights significant changes in dislocation distribution at both micro and nanoscale after compression deformation. A TEM lamella was extracted parallel to the compression direction and perpendicular to the slip bands, as marked by the green arrows in Figure 9i, enabling a direct comparison of dislocation density and distribution with the post-deformation micro-pillar results in Figure 6a.

As shown in Figure 9i, microcracks propagate along the <001> axis before shifting to a 45° angle at the slip bands. This crack propagation pattern resembles that observed in micro-pillar compression tests, although no crack shift was noted in the micro-pillar results (Figure 6b). The presence of both 0° slip lines and the 45° slip lines, as shown in Figure 9j, indicates multi-slip behavior along the {110} planes. Furthermore, the 45° slip lines form a cross-hatched angle of approximately 100°, marked by the black dashed lines in Figure 9j. This exceeds the ~90° cross-hatched angle observed in the 2% strain sample in Figures 8h-k. Importantly, a microcrack initiates at the bottom of the sample along the <001> direction and shifts direction upon intersecting the 0° slip bands, potentially influenced by friction between the indenter and the sample (Zuo and Dienes, 2005). In Figure 9k, after the sample undergoes 15% compression strain, inter-parallel dislocation arrays form within the sample, oriented nearly perpendicular to the surface and parallel to the (001) plane. These dislocation lines migrate toward the surface, resulting in the formation of parallel slip bands, labeled by the green arrows in Figure 9i. The Burger vector associated with these dislocation arrays is $\pm a[10\bar{1}]$, as analyzed in Figure S5 and detailed in Table S2. The combination of bulk compression and DIC measurements suggest dislocation plasticity being the primary deformation mechanism, while microcracking could initiate at stress concentrations.

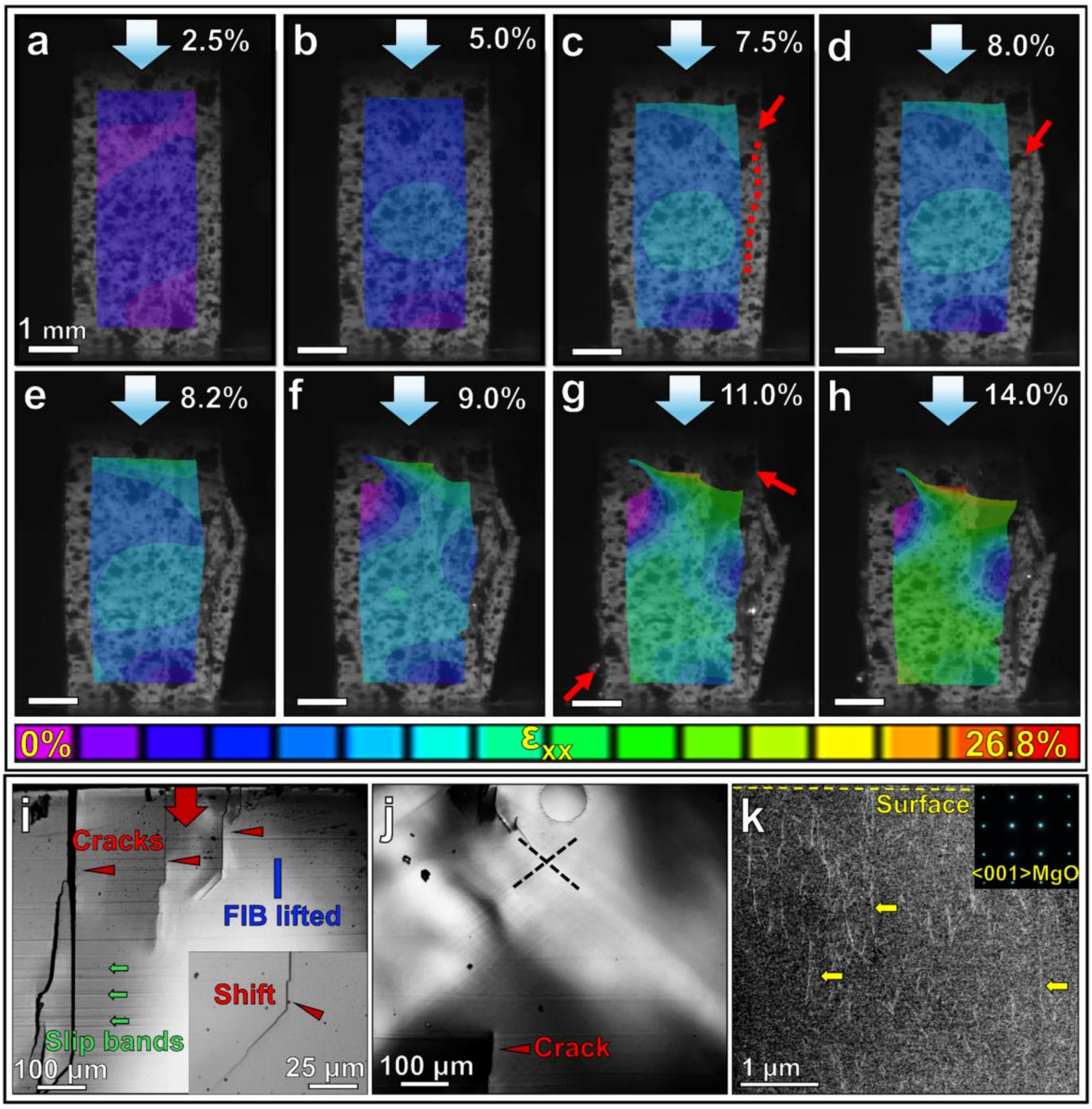

**Figure 9.** 2D distribution of the local strain ($\varepsilon_{xx}$) during compression, and OM images and TEM characterization of the dislocation distribution in the deformed bulk sample after 15% strain. The scale bar in (a) applies to figures b-h. (a) Strain distribution at 2.5% strain with no visible cracks; (b) strain distribution at 5.0% strain, still without noticeable cracks; (c) deformation after 7.5% strain showing the initiation of a vertical crack, marked by the red dashed line, along <001> direction; (d) crack propagation observed at 8.0% strain; (e) further crack growth at 8.2% strain; (f) continued propagation at 9.0% strain; (g) cracks originating from the corner of the specimen, marked by the red arrows, at 11.0% strain; (h) advanced crack propagation observed at 14% strain. (i) OM image of the back side of the deformed bulk sample showing cracks propagating at the top and shifting along {110} plane, as shown in the inset; (j) OM image of the right side revealing cross-hatched slip traces (marked by the black dash lines) and cracks propagating from the bottom; (k) ADF-STEM image along <001> zone axis reveals the dislocation structures after ~15% deformation.

## 4. Discussion

### 4.1. Dislocation multiplication mechanisms

Dislocations and other defects are inherently introduced during the growth of MgO crystals. The critical resolved shear stress for 1/2<110>{110} slip systems in (001) MgO at room temperature is approximately 0.1 GPa (Amodeo et al., 2018). During the Brinell ball scratching process, the calculated indentation pressure is ~1.60 GPa, far exceeding the threshold required for dislocation multiplication in MgO. This level of applied stress readily activates dislocation sources, such as the well-known Frank-Read mechanism, leading to significant dislocation multiplication (Frank and Read, 1950; Messerschmidt and Bartsch, 2003). Room-temperature mechanical processing, such as grinding and scratching, has been shown to induce dislocation multiplication in other oxide ceramics, including $SrTiO_3$ (Fang et al., 2024; Jin et al., 2013).

In LiF crystals, which share similar slip systems with MgO, dislocation multiplication is governed primarily by cross-slip and the formation of jogs on the screw-type dislocations (Johnston and Gilman, 1960). A similar mechanism is observed in MgO, where applied stresses during scratching lead to the formation of new dislocation loops, evidenced by the 45° dislocation lines shown in Figures 1b and 1c. Furthermore, jogs observed in the as-scratched sample (Figure S8) are critical in facilitating dislocation multiplication. Both cross-slip and jog formation play crucial roles in the development of cell walls during plastic deformation. These phenomena are particularly evident in hexagonal-close packed (HCP) materials like magnesium, where they contribute significantly to the generation of point defects, then hinder dislocation motion and lead to the formation of dislocation tangles (Sangiovanni et al., 2022; Tang, 2023). Bi-directional scratching promotes dislocation glide across multiple slip planes, resulting in high-density dislocation tangles and the formation of dislocation cells. Unlike $SrTiO_3$, dislocation multiplication occurs more readily in MgO under identical scratching conditions. For instance, after 1× scratching, the dislocation density in MgO reaches ~ $10^{14}$ $m^{-2}$, whereas in $SrTiO_3$, it is approximately ~ $10^{13}$ $m^{-2}$ (Zhang et al., 2025). This discrepancy arises from MgO's higher pre-existing dislocation density and greater availability of Frank-Read sources introduced during grinding and polishing (Preuß et al., 2024), along with its shorter Burgers vector, all of which enhance dislocation multiplication. Besides, lattice friction stresses are relatively low—65 MPa for edge dislocations and 86 MPa for screw dislocations in MgO (Gaillard et al., 2006), making dislocation glide and multiplication easier.

### 4.2. Micro-pillar compression behavior

#### 4.2.1. Stress drops and strain burst

MgO exhibits pronounced stress drops and strain bursts with few pre-existing dislocation source densities, suggesting stronger dislocation activities (slip events). The dislocation density of the reference MgO micro-pillars is approximately $10^{12}$ $m^{-2}$, and the larger dislocation spacings $L$ beyond 500 nm (Figure 1a) amplify dislocation activities. In micro-pillars with low dislocation densities and small dimensions, dislocations can easily escape to the free surface, resulting in strong slip events characterized by large strain bursts and pronounced serrations in flow stress (Li et al., 2022; Papanikolaou et al., 2018). As the dislocation density increases to ~$10^{14}$ $m^{-2}$, the slip events diminish, leading to smaller stress drops and negligible strain bursts, as shown in Figures 2h and 3h. Upon further increase in the dislocation density (~$10^{15}$ $m^{-2}$), the plastic deformation becomes primarily governed by the dislocation cells (Figure 1c), which limits dislocation escape to the surface and produces minimal stress drops with no strain bursts. Both dislocation density and micro-pillar size influence stress drop and strain burst behavior. For micro-pillars with relatively low dislocation densities ($\rho$ from ~$10^{11}$ to ~$10^{13}$ $m^{-2}$), smaller sizes correlate with more pronounced stress drops and strain bursts, consistent with observations in copper, nickel, and LiF single crystals (Frick et al., 2008; Nadgorny et al., 2008; Zhang et al., 2014).

The distinct characteristics of stress drops and strain bursts are well captured in 3D-DDD simulations, which provide detailed insights into the evolution of dislocation configurations. The stress-strain curves and the snapshots of dislocation microstructure are shown in Figure 10. The stress drops behavior match with the experimental results in Figure 10. For the low initial dislocation density case (Figures 10a-d), plastic deformation is predominantly governed by the operation of limited dislocation sources. As stress increases, typically only one dislocation source is activated. This activated dislocation source rapidly glides across the entire slip plane, causing a significant stress drop. Subsequently, the stress needs to continue increasing to activate the next dislocation source. This results in spatially and temporally intermittent plastic flow, evidenced by discrete slip events, as illustrated in Figures 10b-d, where only the dislocation source indicated by the purple arrow is active.

The dislocation average effective source length ($L = 1/\sqrt{\rho}$) as mention in **Section 3.3** can also interpret the stress drops and strain burst behavior. In the reference sample, the dislocation density is ~$10^{12}$ $m^{-2}$, indicating the average effective source length $L$=1 μm. Hence, the dislocation average effective source length is similar to the dimension of the micro-pillar ($D$ = 0.5 μm). Dislocations are able to move out of the specimen surface with almost no other obstacles in the way of annihilation. The scarcity of dislocation sources induces pronounced discreteness and stochasticity in source activation strengths, resulting in highly intermittent plastic flow characterized by stochastic strain bursts.

For the high initial dislocation density case, strong dislocation interactions promote the formation of junctions and pinning points, resulting in a network of dislocation sources with varying sizes. As shown in Figure 10f, the magnified views (red and blue frames) provide a detailed visualization of local dislocation segment motion at the time instance corresponding to marker f in Figure 10e. Once a dislocation source is initiated, it quickly encounters obstruction from other dislocations, making it difficult to glide across the entire slip plane at one time. This results in smaller stress drop magnitudes. Moreover, due to the abundance of dislocation sources and their relatively small strength differences, another source with comparable strength can be activated almost immediately when a weaker dislocation source fails. This leads to less noticeable stress drops, meaning the stress-strain response appears smoother and more continuous. This is consistence with the experimental results in Figures 2h and 3h. For high density mechanically seeded dislocations (~$10^{15}$ $m^{-2}$, 10× scratched), the average effective source length $L$ decreases to approximately 30 nm. Since the dislocation average effective source length is lower than the size of the micro-pillar, the dislocation source becomes continuous, resulting in smooth stress-strain curves.

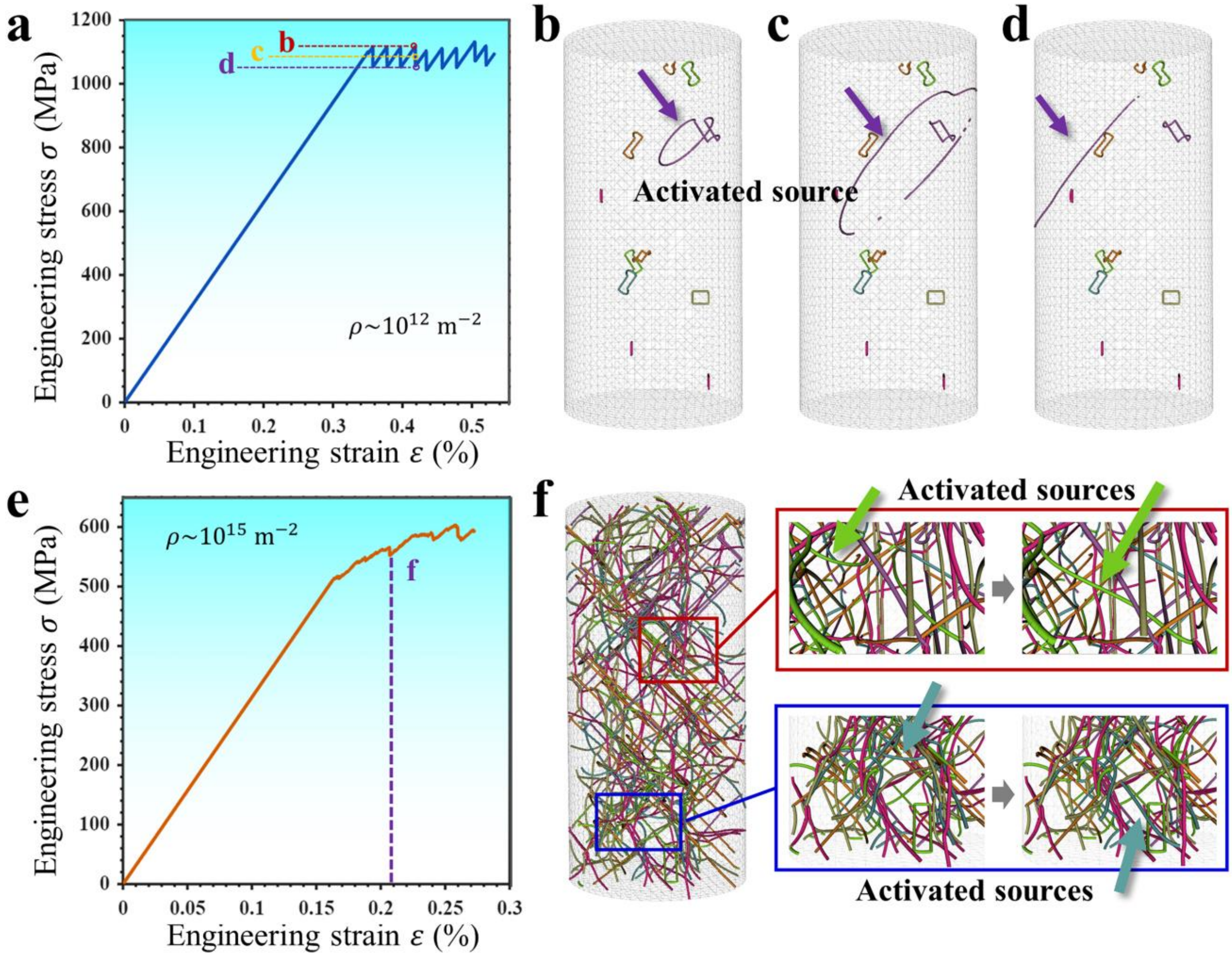

**Figure 10**. Stress-strain curves and corresponding snapshots of dislocation configurations for both low and high initial dislocation densities. The arrows indicate the activated dislocation sources. Stress-strain curves of the micro-pillars with different dislocation densities: (a) $\rho_0$ ~5 × $10^{12}$ $m^{-2}$ and (e) $\rho_0$ ~5 × $10^{14}$ $m^{-2}$. The dislocation configurations in (b–d) and (f) correspond to the markers on the stress-strain curves. Dislocation segments are color-coded according to their Burgers vector type.

### 4.2.2. Yield strength and dislocation work hardening

As summarized in Table 2, the yield strengths of MgO micro-pillars initially decrease and then increase with rising dislocation densities, as depicted in the black dashed line in Figure 11 in **Section 4.2.3**. The yield strength for the 1 μm reference sample with a low dislocation density is ~1 GPa (Figures 2d and 3d). After 1× scratching, the yield strength decreases to ~0.8 GPa as the dislocation density rises to ~$10^{14}$ $m^{-2}$. After 10× scratching, the yield strength increases to nearly 1.5 GPa. This rise reflects dislocation work hardening at the onset of yielding due to increased dislocation density from bi-directional scratching. Moreover, the strain hardening modulus increases with the dislocation density, as displayed in the stress-strain curves in Figures 2h and 2l. The dislocation forest reactions are the main reason for the increased strain hardening modulus (Amodeo et al., 2014). A pronounced dislocation cell structure is observed as the dislocation density reaches ~$10^{15}$ $m^{-2}$ (Figure 1c), and this structure remains highly stable without being noticeably disrupted during deformation. Although it is generally expected that increasing dislocation density enhances dynamic recovery and dislocation annihilation, in the specific high-density regime considered in our work, the samples still exhibit marked strain hardening. Dislocation debris and vacancies generated during cyclic scratching contribute to the enhanced yield strength, consistent with observations in LiF (Gilman and Johnston, 1960). Additionally, dislocation cells formed in the 10× scratched micro-pillars act as barriers to dislocation motion (Kim et al., 2021).

Taylor hardening relationship and limited dislocation source hardening model were employed to predict the change of yield strength modulated by increased dislocations as follows (Cui et al., 2014),

$$\sigma M = \tau_0 + \alpha\mu b\sqrt{\rho} + \frac{k\mu b}{\bar{\lambda}} \tag{4}$$

where $\tau_0$ is the lattice friction stress with magnitude of 86 MPa for screw type dislocations in MgO (Gaillard et al., 2006), $\mu$ is the shear modulus (the value is 133 GPa (Cui et al., 2014)), $b$ is the Burgers vector and its value is 0.3 nm, $k$ is a dimensionless constant and it is taken as 1.0 (Lee and Nix, 2012), $\bar{\lambda}$ is the average effective source length. $M$ is the Schmid factor and equals to 0.5 for the 1/2[01$\bar{1}$](011) and 1/2[10$\bar{1}$](101) slip systems. Based on the experimental data listed in Table 2, the fitted value of $\alpha$ is 0.39. The value $\frac{k}{\bar{\lambda}}$ = 0.004 $nm^{-1}$, hence, $\bar{\lambda}$ is equal to ~250·$k$ nm. For $k$ = 1, D

= 1000 nm, $\bar{\lambda}$ is equal to $\frac{D}{4}$ nm, which indicates the dislocation hardening is related to the dislocation jog mechanism, as displayed in Figure S8. The more significant strain hardening behavior for high initial dislocation density case is also captured by our 3D-DDD simulations. This is found to be contributed by the stronger dislocation interactions and forest hardening. Note that the yield strength calculated by the simulation is lower than that in experiments. This discrepancy is primarily attributed to the absence of effective dislocation cell barriers in simulation. Experimental TEM observations reveal complex self-organized dislocation cells (Figure 1). This is believed to contribute to external strengthening, but it is very challenging to be considered in 3D-DDD due to the very expensive computation cost to consider the very high local dislocation density within the dislocation cell (Hussein and El-Awady, 2016; Wu and Zaiser, 2021). In order to estimate the effect of dislocation cell, we carried out a simplified simulation containing a representative dislocation cell, which had a cell size of approximately 300 nm and the dislocation density on the order of $10^{15}m^{-2}$ inside the cell according to experiments. The short-range slip barrier inside dislocation cell is roughly estimated according to Taylor model. This leads to an addition increase of the yield strength $\Delta\sigma \approx 600$ MPa. Since this calculation underestimates the long-range barrier effect of the dislocation cell, the actual yield strength increase is expected to be higher, therefore, the overall yield strength (>1.1GPa) is expected to agree better with the experiments. In addition, simulation shows that the presence of dislocation cells will further reduce the magnitude of strain bursts. The role of dislocation cell deserves further dedicated investigations.

#### 4.2.3 Size effect

The (001) MgO single crystals exhibit a pronounced size effect in micro-pillar compression tests, both inside (~$10^{15}$ $m^{-2}$) and outside (~$10^{12}$ $m^{-2}$) the scratched tracks (Figure 4). Figure 11b illustrates the size effect observed in this study alongside data from Korte and Clegg (Korte and Clegg, 2011). In their study, the yield strength of the (001) MgO micro-pillar decreases as the diameter increases (green and blue triangles, red rhombic in Figure 11a). The higher yield strength in their work is attributed to the much lower dislocation density (~$10^{10}$ $m^{-2}$), which requires higher stress levels to activate dislocation sources. This results in elevated yield strengths for smaller micro-pillars. In this study, the yield strength also decreases as the diameter of the micro-pillar increases, as displayed in the engineering stress strain curves in Figure 4. However, due to the much higher pre-existing dislocation density (~$10^{15}$ $m^{-2}$), the yield strength decreases from ~1.5 GPa to ~0.5 GPa as the diameter increases from 1 μm to 5 μm. This decrease is likely caused by the activation of additional Frank-Read sources during plastic deformation, which lowers the stress required for dislocation motion (Li et al., 2022; Uchic et al., 2009).

As the micro-pillar diameter increases from 1 μm to 5 μm, the stress-drop behavior observed in the engineering stress-strain curve in Figures 4a and 4b diminishes, resulting in smoother curves. For micro-pillars with the same dislocation density but a smaller diameter, dislocations are easier to escape to the free surface, which leads to stronger dislocation activities and larger stress drops (Nadgorny et al., 2008). Conversely, as the size of the micro-pillar increases, the activation of dislocation sources becomes more continuous, leading to undetectable stress drops. When transitioning from micro-pillar compression to bulk compression, the size effect becomes even more pronounced. The yield strength decreases from ~370 MPa for 5 μm-diameter micro-pillars to ~120 MPa for bulk samples. Similarly, the fracture strain decreases from ~10% in micro-pillars with a 5 μm diameter (Figure 4d) to ~3.6% in bulk samples (Figure 8d).

The yield strengths varied by the pillar diameter with different dislocation densities were analyzed by a power-law fitted in the form of $\frac{\sigma}{\mu} = A\left(\frac{D}{b}\right)^m$ (Dou and Derby, 2009). $\mu$ is the shear modulus (133 GPa), $A$ is a constant, $D$ is the diameter of the pillar, $b$ is the Burgers vector (0.30 nm), and $m$ is the size effect component. As displayed in Figures 11b-d, the fitted size effect components slightly varied with the dislocation density. The micro-pillar size effect components are -0.56, -0.61, and -0.55 for $\rho \sim 10^{10}$ $m^{-2}$, $\rho \sim 10^{12}$ $m^{-2}$, and $\rho \sim 10^{15}$ $m^{-2}$, respectively. The fitted size effect components are slightly higher than the values reported in other ionic crystals, including NaCl, KCl, and LiF, and close to the value in soft fcc metals (–0.6) that exhibited a less pronounced size effect (Korte and Clegg, 2011). The smaller size effect in this study is related to the higher dislocation density, which reaches up to ~$10^{15}$ $m^{-2}$, possessing a dislocation density similar to that of metals.

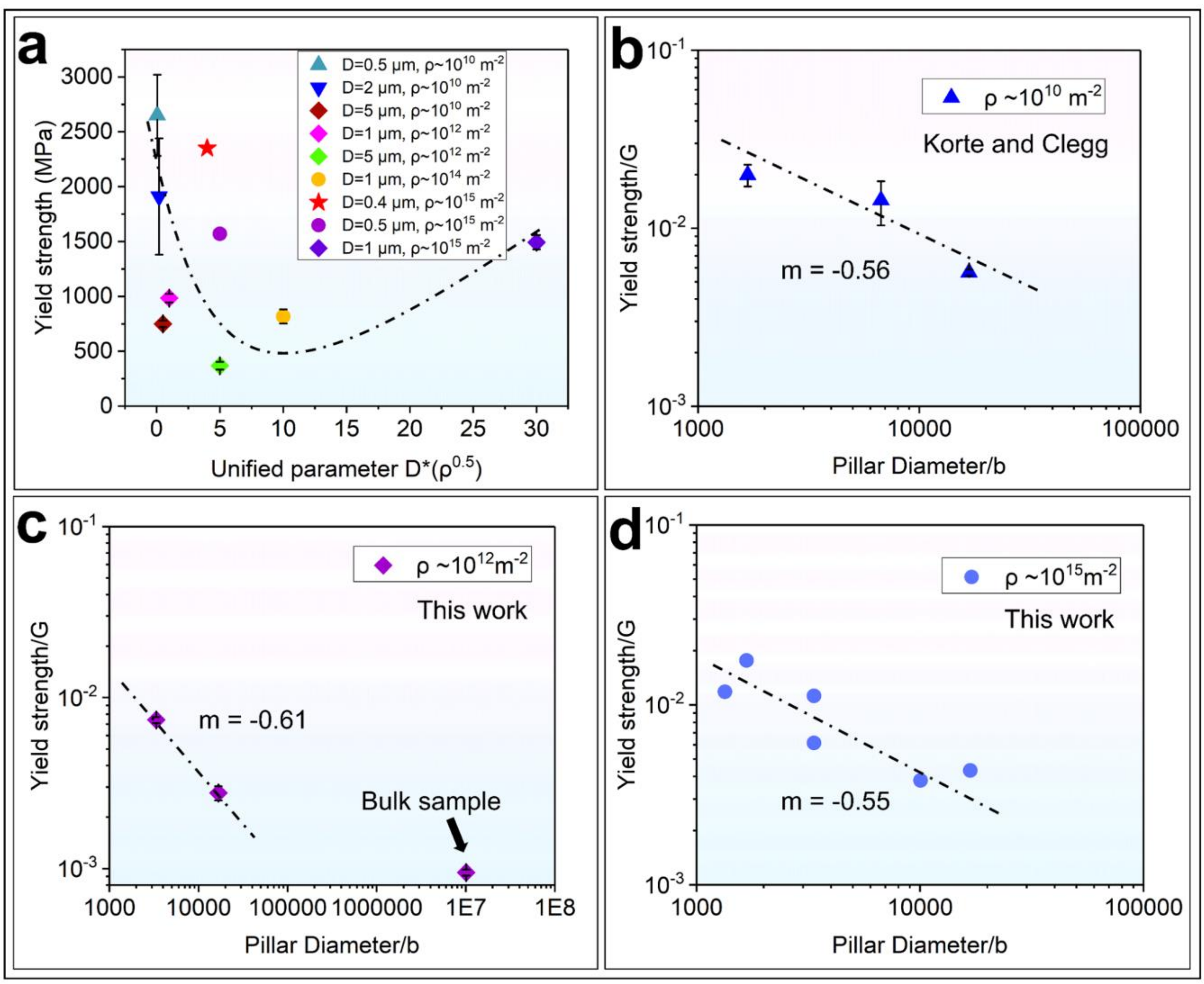

**Figure 11.** Yield strength as the function of dislocation spacing and size effect in (001) MgO single crystals with different dislocation densities. (a) Yield strength as the function of unified parameter $D$ times $1/\sqrt{\rho}$. Data extracted from Korte and Clegg's study (Korte and Clegg, 2011) and the present work. The black dashed line indicates the variation trend of yield strength as modulated by dislocation spacing. Size effect in (001) MgO with different dislocation density (b) $\rho$~$10^{10}$ $m^{-2}$; (c) $\rho$~$10^{12}$ $m^{-2}$; (d) $\rho$~$10^{15}$ $m^{-2}$.

### 4.3. Deformation mechanism in bulk compression

The plastic deformation of MgO single crystals under bulk compression is significantly influenced by pre-existing defect sources, such as Frank-Read sources, point defects, and vacancies introduced during crystal growth. Prior to reaching the yield point, elevated stresses are required to activate these dislocation sources, initiating dislocation motion and multiplication. During the *in situ* bulk compression tests, no slip traces are observed until the yield point is exceeded. This aligns with earlier findings (Stearns et al., 1959; Stokes et al., 1959), which identified the yield stress as the threshold for dislocation activation and multiplication. Once the yield point is reached, dislocation motion and multiplication intensify, as evidenced by the activated {110} slip traces shown in Figure 8b.

As strain increases, slip traces along the {110} planes become more prominent (Figure 8c), indicating a steady rise in dislocation density and activity. These observations correlate well with micro-pillar compression results, where plastic deformation is similarly governed by dislocation dynamics (Figure 3). In MgO, similar to LiF crystals, dislocation multiplication during deformation is primarily influenced by cross-slip and the formation of jogs on screw-type dislocations (Johnston and Gilman, 1960). During bulk compression, cross-slip on the {110} planes significantly contribute to dislocation multiplication. The relatively minor shifts in slip bands observed during the tests reflect this mechanism, demonstrating its role in the plastic deformation of MgO.

### 4.4. Fracture mechanism in micro-pillar and bulk compression

The fracture mechanisms of MgO single crystals vary between micro-pillar and bulk compression tests and are influenced by the pre-existing dislocation sources. For micro-pillar with low dislocation density ($\rho \sim 10^{12}$ m$^{-2}$, Figure 3b), shear deformation occurs along the (011) plane, and cracks form at the edges due to stress concentration. As the dislocation density increases to $\sim 10^{14}$ m$^{-2}$, the micro-pillar deformed to ~40% and exhibited a bending angle of ~26° (Figure 3f). This pronounced bending angle concentrates stress in the bent region, facilitating crack propagation. A similar mechanism has been observed in bending tests of MgO crystals (Kitahara et al., 2016). Conversely, micro-pillars with high dislocation density $\sim 10^{15}$ m$^{-2}$ (10× scratching) deform to 40%, displaying parallel cracks extending vertically from the top surface (Figure 3j). This transition of crack propagation paths highlights the influence of dislocation density on fracture behavior (Figure 3). For the fracture behavior shifts from {110} to {001}/{010} planes, the main reason is the interception of two slip systems from different slip planes, ang this is also reported in materials such as InAs (Howie et al., 2012), silicon (Östlund et al., 2009), and GaAs (Östlund et al., 2011). For the friction between the platen and sample surface, the frictional force at the platen-sample interface is directly proportional to the contact area and influenced by surface roughness. Besides, friction-induced localized stress concentrations may cause crack path deviation. In our study, the surface roughness of the as-polished bulk sample is lower than 2 nm, and the polished $Al_2O_3$ plates offered flat contact interface to minimize the friction force.

In bulk compression, crack initiation primarily occurs along the (010) and (001) planes, often at the intersection of slip planes from two {110} planes, as marked by the red triangle in Figure 8d. The coalescence of dislocations along intersecting slip directions leads to the crack formation on the (010) plane (Keh, 1959). Friction at the indenter-sample interface may also contribute to crack initiation and growth due to the stress concentration on the top of the bulk sample (Huang et al., 2022; Zuo and Dienes, 2005), as shown in Figures 9i and 9j. Elevated frictional stresses and localized strain at the contact surface are evident in the strain field distribution maps in Figures 9e-h. However, the lower pre-existing dislocation density in bulk samples reduces their plasticity. This observation is supported by post-deformation TEM analyses (Figure 9k and Figure S5).

## 5. Conclusion

This study investigates the room-temperature mechanical properties of (001) MgO single crystals with mechanically seeded dislocations, spanning microscale to macroscale deformation. The main conclusions are summarized as follows:

(1) This work demonstrates the modulation of dislocation densities in (001) MgO single crystals, ranging from ~$10^{12}$ $m^{-2}$ to ~$10^{15}$ $m^{-2}$ through room-temperature Brinell cyclic scratching. Dislocations form cells structure in the surface layer at a depth of ~3 μm as the dislocation density reaches ~$10^{15}$ $m^{-2}$.

(2) The tunable, mechanically seeded dislocations enable an exploration of dislocation-driven plastic deformation at the nano-/microscale. Room-temperature micropillar compression tests reveal that as the dislocation density increases from ~$10^{12}$ $m^{-2}$ to ~$10^{14}$ $m^{-2}$ and ~$10^{15}$ $m^{-2}$, the compressive strain rises from 22% ($\rho$ ~$10^{12}$ $m^{-2}$) to ~40% ($\rho$ ~$10^{14}$ $m^{-2}$ and ~$10^{15}$ $m^{-2}$).

(3) The impact of dislocations on the yield strength as well as the work hardening behavior during pillar compression has been corroborated with DDD simulation, confirming the key role of both the dislocation density as well as dislocation structure.

(4) Bulk-scale plasticity investigations combined with DIC highlight a crack propagation mechanism driven by the stress concentration between the contact surfaces and intersection of different slip planes, aligns with observations in micropillar tests.

(5) This study advances the understanding of dislocation-mediated plasticity in oxide ceramics at room temperature, bridging the gap between multiscale deformation and fracture behaviors, offering a basis for future research on dislocation dynamics in ceramics. We also acknowledge our limitations on gradient and uniform dislocation mediated mechanical properties analysis and the absence of cell barriers in the 3D-DDD simulations.

**Acknowledgement:**

W. Lu acknowledge the support by Shenzhen Science and Technology Program (grant no. JCYJ20230807093416034), the Open Fund of the Microscopy Science and Technology-Songshan Lake Science City (grant no. 202401204), National Natural Science Foundation of China (grant no. 52371110) and Guangdong Basic and Applied Basic Research Foundation (grant no. 2023A1515011510). X. Fang thanks the support by the European Union (ERC Starting Grant, Project MECERDIS, grant no. 101076167). Views and opinions expressed are however those of the authors only and do not necessarily reflect those of the European Union or the European Research Council. Neither the European Union nor the granting authority can be held responsible for them. The authors acknowledge using the facilities at the Southern University of Science and Technology Core Research Facility.

**Author contribution:**

W. Lu and X. Fang conceived the idea, designed the experiments, and supervised the project. J. Zhang performed the experimental tests, collected the data, and wrote the first draft. Z. Li and Y. Cui performed the 3D-DDD simulation and wrote the simulation parts. Y. Zhang, H. Holz, J. Best, and O. Preuß performed part of the experimental tests and collected the data. All authors discussed, interpreted the data, and revised the manuscript.

**Conflict of Interest:** The authors declare no conflict of interest.